\documentclass[tightenlines,pra,preprint,showpacs,showkeys,nofootinbib]{revtex4}
\usepackage{graphicx}  %
\usepackage{bm}  %
\usepackage{amsmath}  %

\newcommand{\beq}{\begin{equation}}
\newcommand{\eeq}{\end{equation}}
\newcommand{\bqa}{\begin{eqnarray}}
\newcommand{\eqa}{\end{eqnarray}}

\def\mqo2{{\!\!\!}}

\begin{document}

\title{Condensates of Strongly-interacting Atoms \\
and Dynamically Generated Dimers}

\author{Eric Braaten and Dongqing Zhang}
%\email{braaten@mps.ohio-state.edu}
\affiliation{Department of Physics,
         The Ohio State University, Columbus, OH\ 43210, USA}

\date{\today}
%\date{March 13, 2003}

\begin{abstract}

In a system of atoms with large positive scattering length,
weakly-bound diatomic molecules (dimers) are generated dynamically
by the strong interactions between the atoms.
If the atoms are modeled by
a quantum field theory
with an atom field only,
condensates of dimers
cannot be described by the mean-field approximation
because there is no field associated with
the dimers.
We develop a method for describing dimer condensates
in such a model based on
the one-particle-irreducible (1PI) effective action.
We construct an equivalent 1PI effective action
that depends not only on the classical atom field
but also on a classical dimer field.
The method is illustrated by applying it to
the many-body behavior of bosonic atoms with large scattering length
at zero temperature using an approximation in which
the 2-atom amplitude is treated exactly but irreducible
$N$-atom amplitudes for $N \ge 3$ are neglected.
The two 1PI effective actions give identical results
for the atom superfluid phase, but the one with
a classical dimer field is much more convenient
for describing the dimer superfluid phase.
The results are also compared with previous work
on the Bose gas near a Feshbach resonance.

\end{abstract}

\smallskip
 \maketitle

\section{Introduction}

The development of the technology for
cooling atoms to ultralow temperatures
led to the discovery of Bose-Einstein condensation
in dilute atomic gases~\cite{Anderson1995a,Davis1995b,Bradley1995a}.
It opened up the possibility for experimental study
of Bose gases.
By controlling the scattering length of
the atoms using Feshbach resonances,
it is possible to study strongly-interacting
Bose gases~\cite{Inouye98,Courteille98,Roberts98}.
Among the interesting phenomena
that have been observed is
atom-molecule coherence~\cite{Donley-nature2002-417,Cla02,Rb-85-data},
which is characterized by oscillations whose
frequency coincides with the binding energy
of a diatomic molecule.

There have been several studies
of a Bose gas of atoms near a Feshbach resonance.
Radzihovsky, Park, and Weichman~\cite{RPW}
and Romans, Duine, Sachdev, and Stoof~\cite{RDSS}
predicted that in addition to the normal phase
and an atomic superfluid phase,
there is a molecular superfluid phase.
In the atomic superfluid phase,
there is a Bose-Einstein condensate of atoms
and there may also be a Bose-Einstein condensate
of diatomic molecules (dimer condensate).
In the molecular superfluid phase,
there is a dimer condensate
but no atom condensate.
Lee and Lee used the renormalization group
to study the nature of the phase transition
between the atomic superfluid
and the molecular superfluid~\cite{LL}.
Basu and Mueller studied
various instabilities of the superfluid phases~\cite{BM}.
The studies of Refs.~\cite{RPW} and \cite{RDSS}
indicate that
at zero temperature, there is a quantum
phase transition between the atomic superfluid
and the molecular superfluid.
The relevant interaction parameter is
the detuning energy of the molecule
or, equivalently, the scattering length $a$ of
the atoms. The quantum phase transition occurs at
a positive critical value $a_c$ of the scattering length
that depends on the number density.

A Feshbach resonance allows
the scattering length $a$
to be tuned arbitrarily large.
When the scattering length $a$ is large
compared to the range (and all other length scales
set by the interactions), identical bosons have universal
few-body properties that are determined by $a$
and governed by a discrete scaling symmetry~\cite{Braaten:2004rn}.
Efimov showed in 1970~\cite{Efimov:1970}
that in the unitary limit $a = \pm \infty$,
there are
infinitely many arbitrarily shallow 3-body bound states
(called Efimov trimers).
Their binding energies have a geometric spectrum:
\begin{equation}
E_T ^{(n)} = {22.7}^{-2(n - n_*)}
\hbar ^2 \kappa_* ^2 / m \ ,
\label{Efimov-spec}
\end{equation}
where $m$ is the mass of the bosons
and $\kappa_*$ is the binding wavenumber
of the Efimov trimer labelled by $n_*$.
We will refer to $\kappa_*$
as the Efimov parameter.
The spectrum in Eq.~(\ref{Efimov-spec})
is consistent with a discrete scaling symmetry
with discrete scaling factor
$e^{\pi / s_0} \approx 22.7$,
where $s_0 \approx 1.00624$ is a transcendental number.
Efimov showed that the discrete
scaling symmetry is also relevant
when $a$ is large but not infinite~\cite{Efimov:1971,Efimov:1979}.
For example, few-body observables involving only
particles with zero momentum
must have the form $a^p \ f(s_0 \log (a \kappa_*))$,
where $p$ is
the power required by dimensional analysis
and $f(\theta)$ is a dimensionless function with
period $2 \pi$.
We will refer to such log-periodic behavior
as Efimov physics.

In the previous analyses of the strongly-interacting
Bose gas in Refs.~\cite{RPW,RDSS,LL,BM},
Efimov physics
has been completely ignored.
An obvious qualitative
consequence of Efimov physics
is the possible existence of
a trimer superfluid phase
in which there is a Bose-Einstein condensate
of Efimov trimers (trimer condensate),
but no atom condensate or dimer condensate.
The methods used in
Refs.~\cite{RPW,RDSS,LL,BM}
would be incapable of describing
a trimer superfluid phase.
In these analyses, the model
used as a starting point
was a quantum field theory
with quantum fields
$\hat \psi_A ({\bf r}, t)$ and $\hat \psi_M ({\bf r}, t)$
that annihilate an atom and a molecule, respectively.
Most of the quantitative analyses
in Refs.~\cite{RPW,RDSS}
were carried out using the mean-field approximation.
The order parameters associated with
the atomic superfluid
and molecular superfluid phases
were the mean fields
$\langle \hat \psi_A \rangle$ and
$\langle \hat \psi_M \rangle$,
respectively.
In the atomic superfluid phase,
$\langle \hat \psi_A \rangle \neq 0$ and
$\langle \hat \psi_M \rangle \neq 0$,
while in the molecular superfluid phase,
$\langle \hat \psi_M \rangle \neq 0$ and
$\langle \hat \psi_A \rangle = 0$.
If the scattering length
in this model is tuned to be sufficiently large,
Efimov trimers will be generated dynamically
by the strong interactions between the atoms.
However the mean-field approximation
cannot reveal the existence of a trimer
superfluid phase,
because there is
no quantum field whose expectation value
can indicate the presence of
a condensate of Efimov trimers.

There is an analogous problem
with the dimer superfluid phase in
a quantum field theory
with only an atom field $\hat \psi_A ({\bf r}, t)$.
If the scattering length $a$ is positive
and sufficiently large,
the dimer is generated dynamically
by the strong interactions between the atoms.
The mean field $\langle \hat \psi_A \rangle$
provides an order parameter
for the atom superfluid phase.
However the mean-field approximation cannot reveal
the existence of a dimer superfluid phase,
because there is no quantum field
whose expectation value
can indicate the presence of
a condensate of dimers.

In this paper, we show
how a dimer superfluid phase
can be studied in a quantum field theory
in which the only quantum field
is the atom field $\hat \psi_A ({\bf r}, t)$.
Our basic tool is the
one-particle-irreducible (1PI)
effective action $\Gamma [A]$,
which is a functional of
a classical atom field $A ({\bf r}, t)$.
This effective action encodes
complete information about
the quantum system.
The variational equations
for the 1PI effective action
provide a straightforward description
of states containing an atom condensate
or a mixture of an atom condensate and
a dimer condensate.
They also contain complete information
about states containing a dimer condensate
but no atom condensate,
but this information is encoded
in a more subtle form.
We show how one can construct an equivalent
effective action $\Gamma [A, D]$
that is also a functional of
a classical dimer field $D ({\bf r}, t)$.
This effective action
provides a straightforward description of states containing
a dimer condensate but no atom condensate.

In section II, we introduce
the 1PI effective action
for a quantum field theory.
We determine the 1-body and 2-body terms
$\Gamma_{1+2} [A]$
in the 1PI effective action
for the zero-range model.
In this model, the only interaction
is a contact interaction
that gives the large scattering length $a$.
Any other model with
a large scattering length should reduce to
the zero-range model
when $|a|$ is sufficiently large.
In the case $a > 0$,
we construct the equivalent terms $\Gamma_{1+2} [A,D]$
in an effective action
with a classical dimer field.
In section III, we consider the
approximation defined by the truncated 1PI
effective actions $\Gamma_{1+2} [A]$
and $\Gamma_{1+2} [A, D]$.
This truncated 1PI approximation corresponds to treating
2-atom quantum effects exactly,
but ignoring irreducible 3-atom
and higher $N$-atom effects.
To demonstrate the limitations
of this approximation,
we use it to calculate
the atom-dimer scattering cross-section.
The results are
compared with those from exact solutions
of the 3-body problem, which depend on
the Efimov parameter $\kappa_*$.
In section IV and V, we demonstrate the equivalence of
the truncated 1PI effective actions by using them
to study the strongly-interacting
Bose gas with total atom number density $n$
at zero temperature.
In section IV,
we study the static homogeneous states.
There is an atom superfluid phase
and a dimer superfluid phase
separated by quantum phase transitions
at $a=0$ and $a = (32 \pi n)^{-1/3}$.
In section V,
we study the quasiparticles associated with fluctuations around
the static homogeneous states.
The two truncated 1PI effective actions
give identical results for the atom superfluid phase,
but $\Gamma_{1+2} [A, D]$
is much more convenient for describing
the dimer superfluid phase.
In section VI,
we summarize our results and
discuss their implications
for the Bose gas near a Feshbach resonance.

\section{1PI Effective Action }
\label{ch2}

In this section,
we introduce the 1PI effective action for
a quantum field theory of bosons.
If there is a single atom quantum field,
the 1PI effective action $\Gamma [A]$ is
a functional of a classical atom field $A$
that encodes complete information on quantum effects.
The one-body and two-body terms $\Gamma_{1+2} [A]$
are given exactly for the zero-range model.
For the case $a>0$,
we construct
an equivalent functional $\Gamma_{1+2} [A,D]$
of both the classical atom field $A$
and a classical dimer field $D$.
We also show how thermodynamic properties
of a many-body system
can be calculated
using these 1PI effective actions.

\subsection{Quantum field theory}

The simplest quantum mechanical model that can
describe atoms with a large scattering length $a$
is the {\it zero-range model}.
In the 2-body sector, the zero-range model can be defined by
specifying
the $T$-matrix element for atom-atom scattering.
If the wavevectors of the two atoms in the initial
and final states are $\pm {\vec k}$ and $\pm {\vec k '}$
with $|{\vec k}| = |{\vec k'}| = k$,
the $T$-matrix element is
\begin{equation}
{\cal T} (k) = - \frac{8 \pi \hbar}{m} \frac{a}{1 + i a k} \ ,
\label{T-mat}
\end{equation}
%
%[dz]
where $a$ is the
scattering length.
Since the $T$-matrix element is independent of
the scattering angle,
the zero-range model has $S$-wave
scattering only.
The zero-range model plays
a special role as a minimal model for atoms
with large scattering length.
In any model of atoms whose scattering length
$a$ is large compared to the range,
the observables at energies near the
scattering threshold for atoms
must approach those of the zero-range model as
$|a|$ increases.

The zero-range model can be formulated as a local
quantum field theory.
The quantum field operator satisfies
equal-time commutation relations:
\begin{subequations}
\begin{eqnarray}
\hat \psi ^{\dagger} ({\bf r}, t) \hat \psi ({\bf r}', t) -
\hat \psi ({\bf r}', t) \hat \psi ^{\dagger} ({\bf r}, t)
& = & \delta ({\bf r} - {\bf r}') ,
\\
\hat \psi ({\bf r}, t) \hat \psi ({\bf r}', t) -
\hat \psi ({\bf r}', t) \hat \psi ({\bf r}, t)
& = & 0 .
\end{eqnarray}
\label{commu:boson}
\end{subequations}
%
%[dz,eb]
%
The Hamiltonian $H = \int d^3 r {\cal H}$
is the integral
of a Hamiltonian density:
\begin{equation}
{\cal H} = \hat \psi ^\dagger
\left(
- \frac{\hbar ^2} {2m} \nabla ^2
\right)
\hat \psi \,
+ \,
\frac{1}{4} \lambda(\Lambda)
\hat \psi^\dagger \hat \psi^\dagger
\hat \psi \hat \psi .
\label{H-zrm}
\end{equation}
The time evolution equation is
a partial differential equation:
\begin{equation}
\left(
i \hbar \frac{\partial} {\partial t}
+ \frac{\hbar ^2} {2m} \nabla ^2
\right)
\hat \psi ({\bf r},t) \,
+ \,
\frac{1}{2} \lambda(\Lambda)
\hat \psi ^{\dagger} ({\bf r},t)
\hat \psi ({\bf r},t)
\hat \psi ({\bf r},t) = 0 .
\label{local}
\end{equation}
The product of operators
at the same point
in the interaction term is singular, but it can be
made well-defined by imposing an {\it ultraviolet cutoff}
that restricts the wavevectors
${\bf k}$ in the Fourier expansion of
$\hat \psi ({\bf r}, t)$ to satisfy
$|{\bf k}| < \Lambda $.
To have non-trivial scattering
in the limit $\Lambda \to \infty$,
the bare coupling constant
$\lambda(\Lambda)$
must depend on the ultraviolet cutoff $\Lambda$:
\begin{equation}
\lambda(\Lambda) =
\frac{8 \pi \hbar} {m}
\frac{a} {1 - 2 a \Lambda / \pi} .
\label{run-coupling}
\end{equation}
%
%[dz]
%
Given this coupling constant,
the $T$-matrix element reduces to Eq.~(\ref{T-mat})
in the limit $\Lambda \to \infty.$

The model defined by the Hamiltonian density
in Eq.~(\ref{H-zrm})
is renormalizable in the 2-atom sector.
This means that all 2-atom observables
are well-defined functions of $a$ in the limit
$\Lambda \to \infty$.
For example, the model implies
the existence of a dimer with binding energy
$E_D = \hbar^2 / (m a^2)$.
The model is not renormalizable
in the 3-atom sector.
To make it renormalizable,
it is necessary
to add a 3-body interaction term
$(\hat \psi ^3) ^ \dagger \hat \psi ^3$
to the Hamiltonian density~\cite{BHK99,BHK99b}.
Its coefficient $\lambda_3 (\Lambda)$
can be tuned as a function of $\Lambda$
so that the Efimov parameter $\kappa _*$
has the desired value in the limit
$\Lambda \to \infty$.
All other 3-atom observables
are well-defined functions of
$a$ and $\kappa _*$ in this limit.
There is numerical evidence
that this 3-body interaction
is also sufficient to
make the model renormalizable
in the 4-atom sector~\cite{Platter:2004qn}.
If this is true, it is plausible
that the zero-range model with a 3-body interaction
is also renormalizable in
the $N$-atom sector for all $N$.

The model that was taken as
the starting point
in Refs.~\cite{RPW,RDSS,LL,BM}
was a local quantum field theory
with an atom field $\hat \psi_A ({\bf r}, t)$
and a molecule field $\hat \psi_M ({\bf r}, t)$.
The Hamiltonian density is
\begin{eqnarray}
{\cal H} &=& \hat \psi_A ^\dagger
\left(
- \frac{\hbar ^2} {2m} \nabla ^2
\right)
\hat \psi_A
\, + \,
\hat \psi_M ^\dagger
\left(
- \frac{\hbar ^2} {4m} \nabla ^2
+ \nu_M (\Lambda)
\right)
\hat \psi_M
\nonumber
\\
& + &
\frac{1}{4}
\lambda_{AA} (\Lambda) \,
\hat \psi_A ^\dagger \hat \psi_A ^\dagger
\hat \psi_A \hat \psi_A
\, + \,
\frac{1}{2}
g_{AM} (\Lambda) \,
\left(
\hat \psi_A ^\dagger \hat \psi_A ^\dagger \hat \psi_M +
\hat \psi_M ^\dagger \hat \psi_A \hat \psi_A
\right)
\nonumber
\\
& + &
\lambda_{AM} (\Lambda) \,
\hat \psi_M ^\dagger \hat \psi_A ^\dagger
\hat \psi_A \hat \psi_M
\, + \,
\frac{1}{4}
\lambda_{MM} (\Lambda) \,
\hat \psi_M ^\dagger \hat \psi_M ^\dagger
\hat \psi_M \hat \psi_M .
\label{Res-AD}
\end{eqnarray}
This model is renormalizable
in the 2-atom sector
where the only parameters are
$\nu_M (\Lambda)$,
$\lambda_{AA} (\Lambda)$
and $g_{AM} (\Lambda)$~\cite{Holland02}.
These parameters
can be tuned as functions of
the ultraviolet cutoff $\Lambda$
so that the T-matrix element
for atom-atom scattering in the limit
$\Lambda \to \infty$ is
\begin{equation}
{\cal T} (k) = - \left(
\lambda + \frac{g^2}
{\hbar^2 k^2 /m - \nu}
\right) ^{-1},
\label{T-res}
\end{equation}
where $\lambda$, $g$, and $\nu$
are renormalized parameters
that do not depend on the ultraviolet cutoff.
The scattering length is
\begin{equation}
a = \frac{m} {8 \pi \hbar} \left(
\lambda - \frac{g^2} {\nu}
\right) .
\label{a-res-0}
\end{equation}
It is not known whether
the model defined by the Hamiltonian density
in Eq.~(\ref{Res-AD})
is renormalizable
in the 3-atom or higher $N$-atom sector.
The scattering length in Eq.~(\ref{a-res-0})
can be made
arbitrarily large by tuning to the resonance
$\nu \to 0$.
When this scattering length
is sufficiently large,
observables near the scattering threshold
for atoms must coincide with
those of the zero-range model.
The scattering length must be large
compared to
the background scattering length
$m \lambda / (8 \pi \hbar)$
and also large compared to
the effective range at the resonance,
which is
$-16 \pi \hbar^3 / (m^2 g^2)$.

\subsection{1PI effective action}

The {\it one-particle-irreducible} (1PI) effective action
is a powerful tool
for studying quantum field theories.
The 1PI effective action is introduced in
any modern textbook on quantum field theory,
such as Ref.~\cite{Peskin1995}.
In a quantum field theory with only bosonic fields,
the 1PI effective action is a functional of
a set of classical fields,
one for each quantum field.
The classical field
associated with a quantum field
$\hat \psi ({\bf r}, t)$
that annihilates an atom
is a complex-valued field
that we will denote by $A({\bf r}, t)$.
We will refer to the field $A({\bf r}, t)$
as the {\it classical atom field}.
The 1PI effective action
is a functional of $A({\bf r}, t)$
that we denote by $\Gamma [A]$.
Although $A({\bf r}, t)$ is a classical field,
the functional $\Gamma[A]$ encodes complete information
about the quantum system.

The 1PI effective action
is the generator of 1PI Green functions.
The term in $\Gamma[A]$
that is $N^{\rm th}$ order in both $A$ and $A^*$
gives the 1PI $N$-body Green function.
The 1PI $N$-body Green functions for
$n=1,2, \cdots, N$ are the building blocks
for the $N$-body Green function,
which determines the amplitude for
any $N$-atom quantum state
to evolve into any other $N$-atom quantum state.
The $N$-body Green function can be expressed diagrammatically
as the sum of all possible Feynman diagrams
with $N$ lines coming in, $N$ lines going out,
and arbitrarily many loops.
Using the 1PI effective action formalism,
the connected $N$-body Green function can be expressed more
compactly as a sum of tree diagrams
(i.e. diagrams with no loops)
whose vertices are 1PI $N$-body Green functions
with $n = 2, 3, \ldots, N$.
The 1PI Green functions can be obtained from
variational derivatives of $\Gamma[A]$.
Instead of treating the real and imaginary
parts of $A({\bf r},t)$ as independent variables,
it is more convenient to treat $A({\bf r},t)$
and its complex conjugate $A^* ({\bf r},t)$
as independent.
%% insert 5.0
The $N$-body 1PI Green function can be obtained from
$\Gamma[A]$ by taking $N$ variational derivatives
with respect to $A$
and $N$ variational derivatives
with respect to $A^*$
and then setting $A({\bf r},t) = 0$.

The 1PI effective action $\Gamma[A]$ also encodes
information about the quantum field theory in the presence
of an external source $J ({\bf r},t)$.
The zero-range model in the presence of
an external source is defined by replacing
the zero on the
right side of the time evolution equation
in Eq.~(\ref{local}) by $J ({\bf r}, t)$.
The source is assumed to vanish in the infinite past:
$J ({\bf r}, t) \to 0$ as $t \to - \infty$.
If the system is in the vacuum state in the infinite past,
the state will evolve under the influence of the external source.
In the vacuum state, the expectation value of the quantum field vanishes:
$\langle \hat \psi ({\bf r}, t) \rangle = 0$.
In the presence of an external
source $J({\bf r},t)$, the expectation
value of the quantum field can be nonzero and
it defines a classical field $A({\bf r},t)$:
\begin{eqnarray}
\langle \hat \psi ({\bf r}, t) \rangle_J
= A({\bf r}, t).
\label{Adef}
\end{eqnarray}
%
%[dz,eb]
The subscript $J$ on the angular brackets
in Eq.~(\ref{Adef})
indicates that the expectation value
is evaluated
in the state that evolves from
the vacuum under the influence of the
external source $J ({\bf r}, t)$.
An important basic property of
the 1PI effective action is that
the first variational derivative of
$\Gamma[A]$ with respect to $A^*({\bf r},t)$
gives the source $J({\bf r},t)$ for which
Eq.~(\ref{Adef}) is satisfied:
\begin{eqnarray}
{\delta \phantom{A^* ({\bf r}, t)}
\over
\delta A^* ({\bf r}, t)}
\Gamma[A]
= J({\bf r}, t) .
\label{Jdef}
\end{eqnarray}
%
%[eb,dz]
Higher variational derivatives with respect to
$A({\bf r}, t)$ and $A^* ({\bf r}, t)$
give the 1PI Green function for the
quantum field theory in the presence of the
external source $J({\bf r}, t)$.

If a classical field $A({\bf r}, t)$ satisfies
the variational equation
\begin{eqnarray}
{\delta \phantom{A^* ({\bf r}, t)}
\over
\delta A^* ({\bf r}, t)}
\Gamma[A]
= 0,
\label{Jdef0}
\end{eqnarray}
%
%[eb,dz]
Eq.~(\ref{Jdef}) implies that the corresponding source
$J({\bf r},t)$ vanishes. One can then infer from Eq.~(\ref{Adef})
that $A({\bf r},t)$ can be expressed as an expectation value of
$\hat \psi ({\bf r}, t)$
in the absence of source.
This implies that there is a state $| X_A \rangle$
in the quantum field theory
in which the expectation value of the quantum field
operator is $A({\bf r},t)$:
\begin{eqnarray}
\langle X_A | \hat \psi ({\bf r}, t) |X_A \rangle
= A({\bf r}, t).
\label{XA}
\end{eqnarray}
%
%[eb,dz]
If the source $J({\bf r}, t)$ is nonzero
but vanishes for all $t > t_0$, the vacuum state will
have evolved at time $t_0$ into a nontrivial state $| X_A \rangle$
of the quantum field theory.
The classical field $A({\bf r}, t)$
given by Eq.~(\ref{XA}) will then satisfy the variational
equation~(\ref{Jdef0}) for all $t > t_0$.
Thus each nontrivial solution $A({\bf r}, t)$ of
Eq.~(\ref{Jdef0}) corresponds to a nontrivial state $| X_A \rangle$
of the quantum field theory.

The 1PI effective action $\Gamma [A]$ can be used to study the
static homogeneous states of the quantum field theory.
If the variational equation (\ref{Jdef0})
has a constant solution $A({\bf r}, t)= \bar A$,
there must be a static
homogeneous state $| \bar A \rangle$ in the
quantum field theory in which the expectation
value of the quantum field is $\bar A$:
\begin{equation}
\langle \bar A |
\hat \psi ({\bf r}, t)
| \bar A  \rangle
= \bar A .
\end{equation}
The dynamical stability of that static homogeneous state
can be studied
by considering small-amplitude fluctuations
of $A({\bf r}, t)$
around the constant solution $\bar A$,
i.e. by linearizing the
variational equation (\ref{Jdef0}) around $\bar A$.
If there are any modes whose amplitudes
grow exponentially in time,
then the state $| \bar A \rangle$ is unstable
with respect to quantum fluctuations.
If there are no such modes,
then the state $| \bar A \rangle$
is stable with respect to quantum fluctuations.

Since it is formulated in terms of the expectation value
of a quantum field, the 1PI effective action $\Gamma [A]$
bears a superficial
resemblance to a mean-field approximation.
However it is not a mean-field approximation, but an exact
formulation of the quantum field theory.
The mean-field approximation for the
zero-range model is obtained by replacing
the quantum field $\psi ({\bf r},t)$ in
the evolution equation~(\ref{local})
by a classical field $A ({\bf r},t)$
and by taking the limit $\Lambda \to 0$
in the expression for
the bare coupling constant $\lambda (\Lambda)$
in Eq.~(\ref{run-coupling}).
The resulting partial differential equation is
\begin{equation}
\left(
i \hbar \frac{\partial} {\partial t}
+ \frac{\hbar ^2} {2m} \nabla ^2
\right)
A ({\bf r},t) \,
- \,
\frac{4 \pi \hbar a} {m}
A^{*} ({\bf r},t) A^{2} ({\bf r},t)
= 0 .
\label{local-2}
\end{equation}
This equation can be obtained as the variational equation
for the classical action $S[A]$:
\begin{eqnarray}
S[A] = \int dt \int d^3r \,
\left \{ A^*
\left(
i \hbar \frac{\partial} {\partial t}
+ \frac{\hbar ^2} {2m} \nabla ^2
\right) A
- \frac{8 \pi \hbar a} {m}
(A^{*} A)^2
\right \}.
%\label{Gamma1}
\end{eqnarray}
Since we have taken the limit $\Lambda \to 0$,
the mean-field approximation
ignores all effects of quantum fluctuations.
In contrast,
all quantum fluctuations
are taken into account exactly in the 1PI effective action.

\subsection{Few-body terms}

If the $N$-body Green functions are known analytically,
they can be used to deduce the $N$-body terms in the
1PI effective action.
In the zero-range model,
the 1-body and 2-body Green functions
are known analytically,
limited analytic information is known
about the 3-body sector~\cite{Braaten:2004rn},
and almost no analytic information is known
about the 4-body sector.

\begin{figure}
\centerline{ \includegraphics[width=7.5cm] {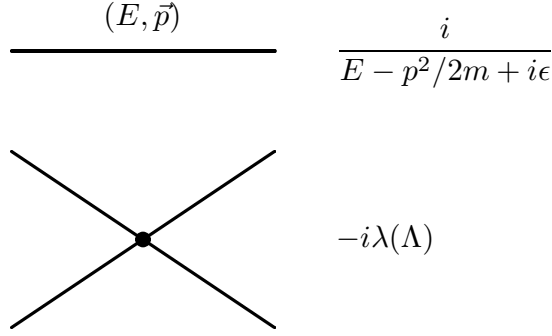}}
\caption{ Feynman rules for the zero-range model.
\label{fig:zero-range}}
\end{figure}

The 1-body term in $\Gamma[A]$ can be deduced from
the propagator for an atom, which is
%$i / (E - p^2 / 2m + i \epsilon)$.
given in Fig.~\ref{fig:zero-range}.
The 1-body term is
\begin{eqnarray}
\Gamma_1[A] = \int dt \int d^3r \,
A^*  \left( i \mbox{${\partial \  \over \partial t}$}
    + \mbox{${1  \over 2}$} \nabla^2 \right) A .
\label{Gamma1}
\end{eqnarray}
%
%[eb,dz]
Here and in much of
the remainder of this paper,
we set $\hbar = m = 1$ for simplicity.
Dimensional analysis can be used to reintroduce factors of
$\hbar$ and $m$ when desired.
The 1-body term in Eq.~(\ref{Gamma1}) implies that
the energy-momentum relation for an isolated atom is
$E = p^2/2m$.

%%%%%%%%%%%%%%%%%%%%%%%%%%%%%%%%%%%%%%%%%%%%%%%%%%%%%%%%%%%%%%%%%%%%%%%%%%%%%%%%%%%%%%%%%%%%%%%
%%%%%%%%%%%%%%%%%%%%%%%%%%%%%%%%%%%%%%%%%%%%%%%%%%%%%%%%%%%%%%%%%%%%%%%%%%%%%%%%%%%%%%%%%%%%%%%

%%%%%%%%%%%%%%%%%%%%%%%%%%%%%%%%%%%%%%%%%%%%%%%%%%%%%%%%%%%%%%%%%%%%%%%%%%%%%%%%%%%%%%%%%%%%%%%
\begin{figure}
\centerline{ \includegraphics[width=12cm] {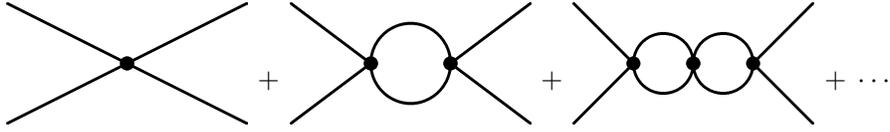}} \caption{
The series of Feynman diagrams in the zero-range model whose sum is
the 2-atom Green function. \label{fig:infty-sum}}
\end{figure}
%%%%%%%%%%%%%%%%%%%%%%%%%%%%%%%%%%%%%%%%%%%%%%%%%%%%%%%%%%%%%%%%%%%%%%%%%%%%%%%%%%%%%%%%%%%%%%%

The 2-body term in $\Gamma[A]$
can be deduced from
the Green function
for atom-atom scattering.
The Feynman rules for the atom-atom interaction vertex
is shown in
Fig.~\ref{fig:zero-range}.
The Green function for atom-atom scattering can be
expressed as the infinite sum of
Feynman diagrams in Fig.~\ref{fig:infty-sum}.
Renormalization is accomplished by using
Eq.~(\ref{run-coupling}) to eliminate
$\lambda (\Lambda)$ in favor of $a$.
The 1PI Green function
depends only on the
total energy $E$ and the total momenta ${\vec p}$
of the two atoms.
Galilean invariance requires
it to depend only on $E - p^2/4m$.
The 1PI Green function is
\begin{eqnarray}
{\cal A}(E, {\vec p}) =
- 8 \pi a \ f(E - p^2/4m),
\label{A2}
\end{eqnarray}
%
%[eb,dz]
where the function $f(x)$ is
\begin{eqnarray}
f (x) = {1 \over 1 - a \sqrt {-x - i \epsilon}}.
\label{f-def}
\end{eqnarray}
%
%[eb,dz]
%
The T-matrix element
for the scattering of two atoms with momenta
$\pm {\bf k}$ is obtained by evaluating
${\cal A}(E, 0)$
at the energy $E = 2(k^2/2)$.
After inserting the factors of $\hbar$ and $m$
required by dimensional analysis,
we recover the $T$-matrix element in Eq.~(\ref{T-mat}).
From the Green function in Eq.~(\ref{A2}), we can deduce
the 2-body term in the 1PI effective action:
\begin{eqnarray}
\Gamma_2[A] =
- 2 \pi a \int dt \int d^3r \, A^{*2}
f \! \left( i \mbox{$\frac{\partial \ }{\partial t}$}
    + \mbox{${1  \over 4}$} \nabla^2 \right)
    A^2 .
\label{Gamma2}
\end{eqnarray}
%
%[eb,dz]

The 1PI Green function in Eq.~(\ref{A2})
contains information about bound states
in the zero-range model.
The poles of the amplitude (\ref{A2})
on the physical sheet of the
complex energy $E$ are the energies of two-body bound states.
The function $f(x)$ in (\ref{f-def}) has a pole
at $x = -1/a^2 + i \epsilon$.
If $a<0$, there are no bound states,
because the pole in $E$ is on the unphysical sheet.
If $a>0$, there is precisely one bound state
and its binding
energy is
$E_D = 1/ {a^2}$.
%as defined in Eq.~(\ref{Edimer}).
We refer to this bound state as the {\it dimer}.
The functional $\Gamma_2[A]$ in Eq.~(\ref{Gamma2})
knows about the existence of the dimer
through the pole at $x =-1/a^2$ of
the function $f(x)$.

%%%%%%%%%%%%%%%%%%%%%%%%%%%%%%%%%%%%%%%%%%%%%%%%%%%%%%%%%%%%%%%%%%%%%%%%%%%%%%%%%%%%%%%%%%%%%%%
\begin{figure}
\centerline{ \includegraphics[width=10cm] {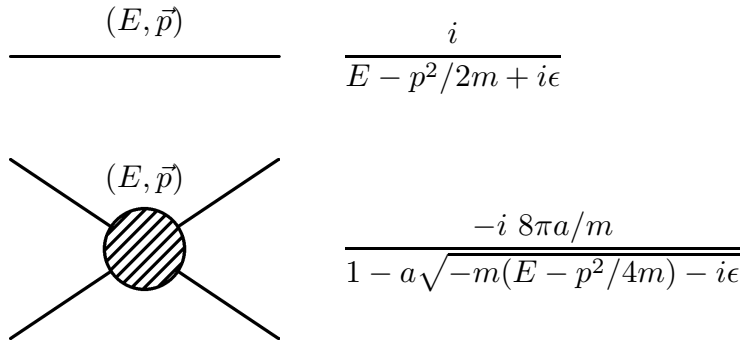} }
\caption{Feynman rules for the truncated 1PI effective action
$\Gamma_{1+2}[A]$. \label{fig:rules-A}}
\end{figure}
%%%%%%%%%%%%%%%%%%%%%%%%%%%%%%%%%%%%%%%%%%%%%%%%%%%%%%%%%%%%%%%%%%%%%%%%%%%%%%%%%%%%%%%%%%%%%%%

The 1PI effective action provides
a compact representation of $N$-body
Green functions as sums of tree diagrams
whose vertices are 1PI Green functions.
The Feynman rules for the 1-body and 2-body terms
in the 1PI effective action
are shown in Fig.~\ref{fig:rules-A}.
The Green function for
atom-atom scattering
is given simply by the vertex in
Fig.~\ref{fig:rules-A},
which depends only on the total
energy $E$ and
total momentum $\vec p$ of the two incoming atoms.
The $T$-matrix element
for the scattering of atoms with momenta $\pm \vec k$ is
obtained by setting $E = 2 (k^2/2m)$
and $\vec p = 0$ in that vertex.
After inserting the factors of $\hbar$ and $m$
required by dimensional analysis,
this reproduces the $T$-matrix element in Eq.~(\ref{T-mat}).

\subsection{Introducing a classical dimer field}
\label{dynamic D}

In the case $a>0$, a pair of atoms
can bind to form the weakly-bound dimer
with binding energy $E_D = 1/a^2$.
The rate for processes
involving dimers can be calculated using
standard diagrammatic methods by using
the Lehmann-Symanzik-Zimmermann (LSZ)
formalism~\cite{Peskin1995}.
This method exploits the fact
that the local composite operator
$\hat \psi ({\bf r}, t) ^2$
has a nonzero amplitude for annihilating a dimer.
The rate for
processes involving dimers
can also be calculated using
the 1PI effective action
by exploiting the fact
that 1PI amplitudes have poles
in the total energy $E$
of any pair of atoms.
For example, the atom-atom scattering vertex
in Fig.~\ref{fig:rules-A}
has a pole at
$E = p^2 / (4 m) - 1/(m a^2)$,
which is the energy of a dimer with
momentum $p$.
One might hope to be able to
calculate such processes more conveniently
using an effective action that is a functional
of a classical dimer field $D({\bf r}, t)$ as well as
the classical atom field $A({\bf r}, t)$.
This motivates the attempt to construct
a functional $\Gamma[A,D]$ that is
equivalent to $\Gamma[A]$.

To construct an effective action $\Gamma[A,D]$
equivalent to $\Gamma[A]$,
we require it to generate the same variational equation
for the classical atom field $A$ after eliminating $D$.
The variational equation obtained by varying
the functional $\Gamma_{1+2} [A]$
%\ref{GammaAnew}
consisting of the sum of
the 1-atom term $\Gamma_1[A]$ in Eq.~(\ref{Gamma1})
and the 2-atom term $\Gamma_2[A]$ in Eq.~(\ref{Gamma2})
is
\begin{equation}
\left( i \mbox{$\frac{\partial \ }{\partial t}$}
    + \mbox{${1  \over 2}$}  \nabla^2 \right) A
= 4 \pi a A^* f \! \left( i \mbox{$\frac{\partial \ }{\partial t}$}
    + \mbox{${1  \over 4}$} \nabla^2 \right)
    A^2,
\label{varyAnew}
\end{equation}
where $f(x)$ is the function defined in Eq.~(\ref{f-def}).
Since $f(x)$ has a pole at $x= -1/a^2$,
the right side of Eq.~(\ref{varyAnew}) includes
a term in which the operator
$\left( i \mbox{$\frac{\partial \ }{\partial t}$}
    + \mbox{$\frac{1}{4}$} \nabla^2
    + 1/a^2
\right) ^{-1}$
acts on $A^2$.
We can formally define a classical dimer field $D ({\bf r},t)$
by the equation
\begin{equation}
D({\bf r},t) = \mbox{$1 \over 2$} g \left( i \mbox{$\frac{\partial \
}{\partial t}$} + \mbox{${1  \over 4}$} \nabla^2 + 1/a^2
\right)^{-1} A^2({\bf r},t),
\label{D-def-new}
\end{equation}
where $g$ is
%a constant determined by the normalization of $D$.
for now an arbitrary constant.
%
% With an explicit dimer field
%as defined in Eq.~(\ref{D-def-new}),
The variational equation (\ref{varyAnew})
can be expressed as a
pair of coupled equations for $A$ and $D$:
\begin{subequations}
\begin{eqnarray}
\left( i \mbox{$\frac{\partial \ }{\partial t}$}
    + \mbox{${1  \over 2}$} \nabla^2 \right) A
&=& { 16 \pi \over a g} A^* D + 4 \pi a A^* \bar f \! \left( i
\mbox{$\frac{\partial \ }{\partial t}$}
    + \mbox{${1  \over 4}$} \nabla^2 \right)
    A^2 ,
\label{varyAD:1new}
\\
\left( i \mbox{$\frac{\partial \ }{\partial t}$}
    + \mbox{${1  \over 4}$} \nabla^2 \right) D &=&
- {1 \over a^2} D + \mbox{$1 \over 2$} g A^2 ,
\label{varyAD:2new}
\end{eqnarray}
\label{varyADnew}
\end{subequations}
where $\bar f(x)$ is a function that is regular at $x = -1/a^2$:
\begin{subequations}
\begin{eqnarray}
\bar f(x) & = & f(x) - {2 \over 1+a^2 x + i \epsilon}
\label{fbar1}
\\
& = & \frac{-1} { 1 + a \sqrt{-x - i \epsilon} } .
\label{fbar2}
\end{eqnarray}
\label{fbar}
\end{subequations}
%
%[dz,eb]
The definition of the classical dimer field in
Eq.~(\ref{D-def-new}) follows from Eq.~(\ref{varyAD:2new}).
Upon using Eq.~(\ref{D-def-new}) to
eliminate $D$ from Eq.~(\ref{varyAD:1new}), we recover
Eq.~(\ref{varyAnew}) for the time evolution
of the classical atom field.

The Eqs.~(\ref{varyADnew}) can be obtained as variational equations
from an effective action $\Gamma_{1+2} [A,D]$
that is a functional of $A({\bf r}, t)$ and $D({\bf r}, t)$.
The constant
$g$ in the definition of $D$ in Eq.~(\ref{D-def-new}) can be
determined by demanding that the integrand of the functional
includes the standard $1$-body term for a dimer field: $ D^* \left(
i \mbox{$\frac{\partial \ }{\partial t}$} + \mbox{${1 \over 4}$}
\nabla^2 \right) D $. The resulting value of $g$ is
\begin{equation}
g = (16 \pi/a)^{1/2}.
\label{g}
\end{equation}
%
%[dz, eb]
The effective action is
\begin{eqnarray}
\Gamma_{1+2} [A,D] &=&
\int dt \int d^3r \left\{ A^*  \left( i
\mbox{${\partial \  \over \partial t}$}
    + \mbox{${1  \over 2}$} \nabla^2 \right) A
+ D^*  \left( i \mbox{$\frac{\partial \ }{\partial t}$}
    + \mbox{${1 \over 4}$} \nabla^2 + 1/a^2 \right) D \right.
\nonumber
\\
&& \hspace{2 cm}
\left. - \mbox{$1 \over 2$} g \left(A^{*2} D + D^*
A^2 \right) - 2 \pi a A^{*2} \bar f \! \left( i
\mbox{$\frac{\partial \ }{\partial t}$}
    + \mbox{${1  \over 4}$} \nabla^2 \right)
    A^2 \right\} .
\label{GammaADnew}
\end{eqnarray}
By varying this functional,
we obtain the pair of coupled
variational equations~(\ref{varyADnew}).

The method used to construct the functional
$\Gamma_{1+2} [A,D]$ in Eq.~(\ref{GammaADnew}),
which includes all 1-atom and 2-atom terms,
could in principle be used to construct
a complete 1PI effective action $\Gamma [A,D]$
that includes all higher $N$-atom terms.
If the $N$-atom Green function
were known analytically,
terms that contain poles
in the total energy of
an incoming or outgoing pair of atoms
would be replaced by
terms with a classical
dimer field $D$ or $D^*$.
The resulting 1PI effective action $\Gamma [A,D]$
for classical atom and dimer fields
would be equivalent to the conventional
1PI effective action $\Gamma [A]$.

%%%%%%%%%%%%%%%%%%%%%%%%%%%%%%%%%%%%%%%%%%%%%%%%%%%%%%%%%%%%%%%%%%%%%%%%%%%%%%%%%%%%%%%%%%%%%%%
\begin{figure}
\centerline{ \includegraphics[width=10cm] {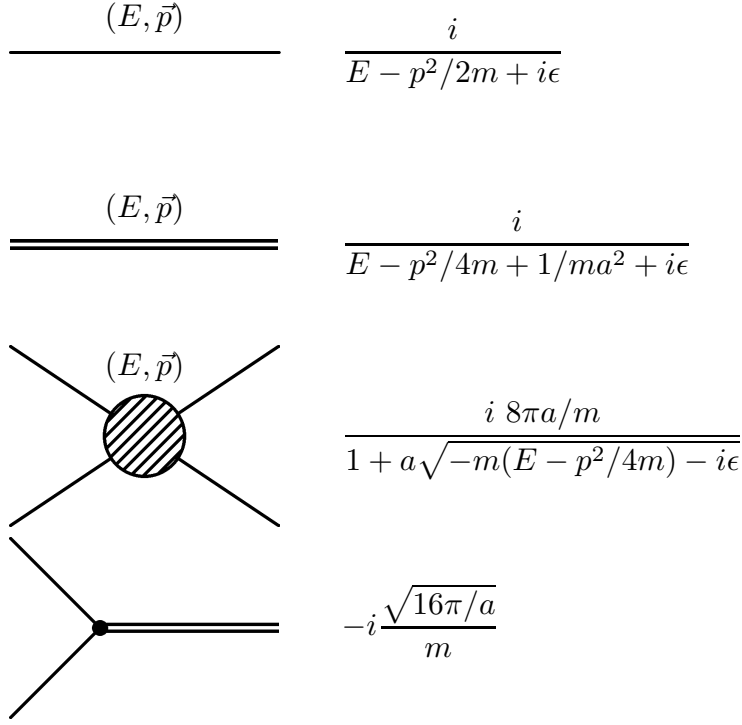} }
\caption{Feynman rules for the truncated 1PI effective action
$\Gamma_{1+2}[A,D]$.
\label{fig:rules-AD}}
\end{figure}
%%%%%%%%%%%%%%%%%%%%%%%%%%%%%%%%%%%%%%%%%%%%%%%%%%%%%%%%%%%%%%%%%%%%%%%%%%%%%%%%%%%%%%%%%%%%%%%

\begin{figure}
\centerline{ \includegraphics[width=10cm] {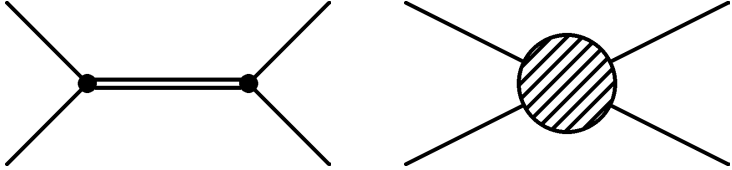} }
\caption{Feynman diagrams for atom-atom scattering from the
truncated 1PI effective action $\Gamma[A,D]$. \label{fig:ADA-scat}}
\end{figure}

%
Using the 1PI effective action $\Gamma[A,D]$,
the Green function for $N$ atoms could be represented
as the sum of tree diagrams with
$N$ incoming atom lines, $N$ outgoing atom lines,
and vertices connected by
either atom lines or dimer lines.
The Feynman rules for the 1-atom and 2-atom terms
in the effective action
$\Gamma_{1+2} [A,D]$
are shown in Fig.~\ref{fig:rules-AD}.
They include
propagators for the atom and the dimer,
an atom-atom scattering vertex,
and a point coupling between two atoms and a dimer.
The Feynman rule for the atom-atom scattering vertex
in Fig.~\ref{fig:rules-AD}
is different from the one
in Fig.~\ref{fig:rules-A}.
The Green function for atom-atom scattering
is the sum of the two tree diagrams in
Fig.~\ref{fig:ADA-scat}.
The $T$-matrix for the scattering of atoms
with momenta $\pm \vec{k}$ is obtained
by setting
$E = 2 (k^2/2m)$ and $\vec p= 0$.
After inserting the factor of $\hbar$
required by dimensional analysis,
this reproduces the $T$-matrix element in Eq.~(\ref{T-mat}).
%Thus we can see that
%the sum of the two tree diagrams in Fig.~\ref{fig:ADA-scat}
%associated with $\Gamma[A,D]$
%is equal to the vertex in Fig.~\ref{fig:rules-A}
%associated with $\Gamma [A]$.

The 1PI effective action $\Gamma[A,D]$
also provides a convenient representation
of amplitudes for processes with dimers
in the initial and final states.
Such an amplitude is given by
the sum of tree diagrams with an
external dimer line for
each of the initial or final dimers.
This amplitude could also be derived from
$\Gamma[A]$ by using the LSZ formalism~\cite{Peskin1995}
which requires finding the residues of poles
in the total energy of pairs
of external atom lines.
The effective action $\Gamma[A,D]$
gives these amplitudes much more directly.

%%%%%%%%%%%%%%%%%%%%%%%%%%%%%%%%%%%%%%%%%%%%%%%%%%%%%%%%%%%%%%%%%%%%%%%%%%%%%%%%%%%%%%%%%%%%%%%
%%%%%%%%%%%%%%%%%%%%%%%%%%%%%%%%%%%%%%%%%%%%%%%%%%%%%%%%%%%%%%%%%%%%%%%%%%%%%%%%%%%%%%%%%%%%%%%

The value of the atom-dimer coupling constant
$g$ in Eq.~(\ref{g})
has been obtained previously
in Ref.~\cite{RDSS}.
The authors expressed the atom-molecule interaction
term in the Hamiltonian density Eq.~(\ref{Res-AD}) as
$\sqrt{Z} g \ (\hat \psi_M ^\dagger \hat \psi_A \hat \psi_A + h.c.)$,
where $\sqrt{Z}$ is the amplitude for a ``dressed molecule"
to be a ``bare molecule".
As $\nu \to 0$,
which implies $a \to \pm \infty$,
the limiting behavior of this amplitude is
$\sqrt{Z} \to 2 \sqrt{\pi} \hbar^2 / (gm \sqrt{a})$.
Setting $\hbar = m =1$,
the limiting behavior of the coefficient of
$\hat \psi_M ^\dagger \hat \psi_A \hat \psi_A$
is $\sqrt{Z} g \to \sqrt{4 \pi /a}$.
This agrees with the coefficient
$\mbox{$\frac{1}{2}$} g$
of the term $D^* A^2$
in the integrand of the effective action functional
in Eq.~(\ref{GammaADnew}).

\subsection{Many-body problem}

The 1PI effective action can
also be applied to a many-body system
containing infinitely many atoms
with a given local number density $n ({\bf r},t)$.
If $\hat {\cal H}$ is the Hamiltonian density operator,
the many-body system can be described
by a quantum field theory
whose Hamiltonian density operator is
$\hat {\cal H} - \mu \ \hat \psi^\dagger \hat \psi$,
where $\mu$ is the chemical potential.
In the $N$-body sector, the only effect of
a constant chemical potential $\mu$
is to shift the total energy by $N \mu$.
If the system is homogeneous,
$\mu$ can be adjusted so that
the number density $n$
has the desired value.
In a nonhomogeneous many-body system,
the function $\mu({\bf r},t)$
can be chosen so that the ground state of the system
has the desired local number density $n({\bf r},t)$.
Position-dependent terms
in $\mu({\bf r},t)$
play the rule of a trapping potential.
In the case of
a time-independent chemical potential $\mu ({\bf r})$,
the 1PI effective action $\Gamma [A; \mu]$
can be obtained from the conventional
1PI effective action $\Gamma [A]$ simply
by replacing the time derivative
$i {\partial \over \partial t}$
with $i {\partial \over \partial t} + \mu$
when it acts on $A$ as in Eq.~(\ref{Gamma1}),
with $i {\partial \over \partial t} + 2 \mu$
when it acts on $A^2$ as in Eq.~(\ref{Gamma2}), etc.
% or on $D$ as in (\ref{GammaADnew})
%
If the chemical potential $\mu ({\bf r})$
is time dependent, obtaining $\Gamma [A; \mu]$
is more difficult because $\mu$
does not commute with the
time derivative $\frac{\partial} {\partial t}$.

The local number density $n({\bf r},t)$
is the expectation value
of the number density operator:
\begin{equation}
n ({\bf r},t) =
\langle \hat \psi^\dagger ({\bf r},t)
\hat \psi ({\bf r},t) \rangle.
\label{ndef}
\end{equation}
%[dz, eb]
It is the conserved density associated with the invariance
of the 1PI effective action $\Gamma[A;\mu]$ under the phase
symmetry $A \to e^{i \alpha} A$.
It can therefore be derived from
the effective action using the
generalization of Noether's prescription to
the case of arbitrarily
high-order time derivatives. Alternatively,
it can be obtained
simply by varying $\Gamma[A;\mu]$ with respect
to the chemical potential:
\begin{equation}
n ({\bf r}, t) =
{\delta \phantom{\mu ({\bf r}, t)}
    \over
\delta \mu ({\bf r}, t)}
\Gamma[A; \mu].
\label{density}
\end{equation}
%
%[eb,dz]

The 1PI effective action $\Gamma[A,D]$ with
a classical dimer field
can also be generalized to
a functional $\Gamma[A,D;\mu]$
that describes the many-body system
with chemical potential $\mu ({\bf r}, t)$.
From the definition of the classical dimer field
in Eq.~(\ref{D-def-new}), we can see that
its behavior under the phase symmetry is
$D \to e^{i 2 \alpha} D$.
In the case of
a time-independent chemical potential $\mu ({\bf r})$,
$\Gamma [A,D; \mu]$
can be obtained from $\Gamma[A,D]$
by replacing the time derivative
$i {\partial \over \partial t}$
with $i {\partial \over \partial t} + \mu$
when it acts on $A$,
% as in Eq.~(\ref{Gamma1}),
with $i {\partial \over \partial t} + 2 \mu$
when it acts on $A^2$ or $D$, etc.
%as in Eq.~(\ref{GammaADnew}).
%
The number density $n({\bf r},t)$
can be determined
by using the fact that it is the conserved density
associated with the phase symmetry.
Thus it can be obtained by
varying $\Gamma[A,D;\mu]$ with respect
to the chemical potential
as in Eq.~(\ref{density}).

The energy density ${\cal E}$ is
the expectation value of the
hamiltonian density operator:
${\cal E} = \langle \hat {\cal H} \rangle$.
The free energy density
${\cal F} = {\cal E} - \mu n$
is the conserved density
associated with the invariance of
the effective action $\Gamma[A; \mu]$
under translations in time:
$A({\bf r},t) \to  A({\bf r},t+ \epsilon)$.
It can therefore be derived
by using Noether's method generalized to
the case of arbitrarily high order time derivatives.
Similarly, if we use the effective action $\Gamma[A,D; \mu]$
with a classical dimer field,
the free energy density
can be determined
by using the fact that
it is the conserved density
associated with translations in time
of $A({\bf r},t)$ and $D({\bf r},t)$.

\section{Truncated 1PI Effective Action}
\label{ch3}

In this section,
we introduce an approximation
in which the 1-body and 2-body terms
in the 1PI effective action
are included exactly,
but 3-body and higher-body terms are neglected.
We refer to this approximation as
the truncated 1PI approximation.
The cross section for atom-dimer scattering
is calculated using
the truncated 1PI approximation and
compared to results from the exact solution of
the 3-body problem.

\subsection{Classical atom field only}

We define the truncated 1PI effective action
for the many-body system by
\begin{eqnarray}
\Gamma _{1+2}[A; \mu] = \int dt \int d^3r \left\{ A^*  \left( i
\mbox{${\partial \  \over \partial t}$} + \mu
    + \mbox{${1  \over 2}$} \nabla^2 \right) A
- 2 \pi a A^{*2} f \! \left( i \mbox{$\frac{\partial \ }{\partial
t}$} + 2 \mu
    + \mbox{${1  \over 4}$} \nabla^2 \right)
    A^2 \right\},
\label{GammaA}
\end{eqnarray}
%
%[eb,dz]
where $f(x)$ is the function defined in Eq.~(\ref{f-def}).
The variational equation
for the classical field $A ({\bf r},t)$ is
\begin{equation}
\left( i \mbox{$\frac{\partial \ }{\partial t}$} + \mu
    + \mbox{${1  \over 2}$}  \nabla^2 \right) A
= 4 \pi a A^* f \! \left( i \mbox{$\frac{\partial \ }{\partial t}$}
+ 2 \mu
    + \mbox{${1  \over 4}$} \nabla^2 \right)
    A^2.
\label{varyA}
\end{equation}
%
%[eb,dz]
Note that this equation is nonlocal in position and time,
because $f$ is a nonpolynomial function of an operator that involves
the derivatives $\frac{\partial}{\partial t}$ and $\nabla$.
A solution of this equation corresponds to a state
containing a Bose-Einstein condensate of atoms
whose common wavefunction is proportional to $A ({\bf r},t)$.

The number density $n ({\bf r},t)$
that follows
from the truncated 1PI effective action
in Eq.~(\ref{GammaA}) is given by the variational derivative
with respect to $\mu ({\bf r},t)$
in Eq.~(\ref{density}).
In the special case of a constant chemical potential $\mu$,
the variational derivative
can be evaluated explicitly:
\begin{eqnarray}
n = A^*  A - 4 \pi a A^{*2} f' \! \left( i \mbox{$\frac{\partial \
}{\partial t}$} + 2 \mu
    + \mbox{${1  \over 4}$} \nabla^2 \right)
    A^2 ,
\label{n-A}
\end{eqnarray}
%
%[eb,dz]
where $f'(x)$ is the derivative of the function defined in
Eq.~(\ref{f-def}):
\begin{eqnarray}
f'(x) = {- a \over 2 \sqrt {-x - i \epsilon} \left( 1 - a \sqrt {-x
- i \epsilon} \right)^2}. \label{fp}
\end{eqnarray}
%
%[eb,dz]

The free energy density ${\cal F} ({\bf r},t)$
that follows from the truncated 1PI effective action in
Eq.~(\ref{GammaA}) can be obtained by applying Noether's theorem
to the time-translation symmetry.
In the special case of a
time-independent classical field $A({\bf r})$, Noether's
prescription reduces to
\begin{eqnarray}
{\cal F} = - A^*  \left( \mu + \mbox{${1  \over 2}$} \nabla^2
\right) A + 2 \pi a A^{*2} f \! \left( 2 \mu
    + \mbox{${1  \over 4}$} \nabla^2 \right)
    A^2 .
\label{F-A}
\end{eqnarray}
%
%[eb,dz]

\subsection{Classical atom and dimer fields}

In the case $a>0$,
we can construct an equivalent truncated 1PI effective action
that is a functional of classical atom and dimer fields:
\begin{eqnarray}
\Gamma_{1+2} [A,D; \mu] &=& \int dt \int d^3r \left\{ A^*  \left( i
\mbox{${\partial \  \over \partial t}$} + \mu
    + \mbox{${1  \over 2}$} \nabla^2 \right) A
+ D^*  \left( i \mbox{$\frac{\partial \ }{\partial t}$} + 2 \mu
    + \mbox{${1 \over 4}$} \nabla^2 + 1/a^2 \right) D \right.
\nonumber
\\
&& \hspace{2 cm} \left. - \mbox{$1 \over 2$} g \left(A^{*2} D + D^*
A^2 \right) - 2 \pi a A^{*2} \bar f \! \left( i
\mbox{$\frac{\partial \ }{\partial t}$} + 2 \mu
    + \mbox{${1  \over 4}$} \nabla^2 \right)
    A^2 \right\} ,
\label{GammaAD}
\end{eqnarray}
%
%[dz,eb]
where $\bar f (x)$ is defined in
Eq.~(\ref{fbar}) and
$g = \sqrt{16 \pi /a}$.
The equations obtained by varying
this functional with respect to $A^* ({\bf r},t)$
%with $A ({\bf r},t)$ fixed
and $D^* ({\bf r},t)$
%with $D ({\bf r},t)$ fixed:
are
\begin{subequations}
\begin{eqnarray}
\left( i \mbox{$\frac{\partial \ }{\partial t}$} + \mu
    + \mbox{${1  \over 2}$} \nabla^2 \right) A
&=& { 16 \pi \over a g} A^* D + 4 \pi a A^* \bar f \! \left( i
\mbox{$\frac{\partial \ }{\partial t}$} + 2 \mu
    + \mbox{${1  \over 4}$} \nabla^2 \right)
    A^2 ,
\label{varyAD:1}
\\
\left( i \mbox{$\frac{\partial \ }{\partial t}$} + 2 \mu
    + \mbox{${1  \over 4}$} \nabla^2 \right) D &=&
- {1 \over a^2} D + \mbox{$1 \over 2$} g A^2 .
\label{varyAD:2}
\end{eqnarray}
\label{varyAD}
\end{subequations}
%
%[dz,eb]
The formal solution of Eq.~(\ref{varyAD:2}) for
the field $D({\bf r}, t)$ is
\begin{equation}
D({\bf r},t) = \mbox{$1 \over 2$} g
\left(
i \mbox{$\frac{\partial \ }{\partial t}$}
+ 2 \mu + \mbox{$\frac{1}{4}$} \nabla^2 + 1/a^2
\right)^{-1} A^2({\bf r},t).
\label{D-def}
\end{equation}
%
%[dz,eb]
%
Upon using this solution to eliminate
$D$ from Eqs.~(\ref{varyAD:1}),
we recover Eq.~(\ref{varyA}) for the
classical atom field.
A solution of Eqs.~(\ref{varyAD})
corresponds to a state containing
a condensate of atoms whose common wavefunction
is proportional to $A({\bf r}, t)$
and a condensate of dimers
whose common center-of-mass wavefunction
is proportional to $D({\bf r}, t)$.

The number density $n({\bf r},t)$ associated with
the truncated 1PI effective action in Eq.~(\ref{GammaAD})
is given by
the variational derivative with respect to $\mu({\bf r},t)$
in Eq.~(\ref{density}).
In the special case of a constant chemical potential $\mu$,
the variational
derivative can be evaluated explicitly:
\begin{eqnarray}
n =& A^*  A + 2 D^*  D - 4 \pi a A^{*2} \bar f' \! \left( i
\mbox{$\frac{\partial \ }{\partial t}$} + 2 \mu + \mbox{${1  \over
4}$} \nabla^2 \right) A^2 ,
\label{n-AD}
\end{eqnarray}
%
%[eb,dz]
where ${\bar f}'(x)$ is the derivative of the function
defined in Eq.~(\ref{fbar}):
\begin{equation}
\bar f' (x) = \frac{-a /2} { ( 1 + a \sqrt{-x - i \epsilon} )^2
\sqrt{-x - i \epsilon} } .
\label{fbar'}
\end{equation}
%
%[dz]

The energy density ${\cal E} ({\bf r},t)$ that
follows from the truncated 1PI effective action in
Eq.~(\ref{GammaAD}) can be obtained by applying Noether's theorem
to the time-translation symmetry.
In the special case of time-independent classical fields $A({\bf r})$ and
$D({\bf r})$, Noether's prescription reduces to
\begin{eqnarray}
{\cal F}  &=& - A^*  \left( \mu
    + \mbox{${1  \over 2}$} \nabla^2 \right) A
- D^*  \left( 2 \mu
    + \mbox{${1 \over 4}$} \nabla^2 + 1/a^2 \right) D
\nonumber
\\
&&+ \mbox{$1 \over 2$} g \left(A^{*2} D + D^* A^2 \right) + 2 \pi a
A^{*2} \bar f \! \left( 2 \mu
    + \mbox{${1  \over 4}$} \nabla^2 \right)
    A^2  .
\label{F-AD}
\end{eqnarray}
%
%[eb,dz]

\subsection{Atom-dimer elastic scattering}
\label{section:AD}

The truncation of the 1PI effective action
to include only 1-atom and 2-atom terms
guarantees that 2-atom effects are treated exactly.
In this approximation, irreducible $N$-atom
amplitude for $N \ge 3$ are neglected.
This does not imply
that higher $N$-atom effects are trivial,
because they can arise from
the iteration of 2-atom effects.
In this subsection, we clarify the meaning of
the truncated 1PI effective action
by using it to calculate
the cross section for atom-dimer elastic scattering.
We will compare the cross section
with results from
exact solutions of the 3-body problem.
In this subsection,
we set $\mu = 0$ and
make the mass $m$ explicit in the equations
instead of setting $m=1$.

%%%%%%%%%%%%%%%%%%%%%%%%%%%%%%%%%%%%%%%%%%%%%%%%%%%%%%%%%%%%%%%%%%%%%%%%%%%%%%%%%%%%%%%%%%%%%%%
\begin{figure}
\centerline{ \includegraphics[width=7cm] {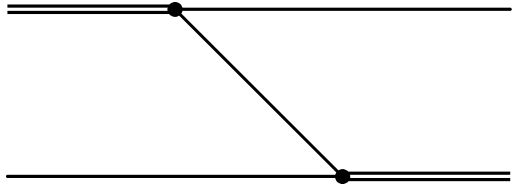} }
\caption{Feynman diagram for elastic atom-dimer scattering from the
truncated 1PI effective action $\Gamma_{1+2}[A, D]$.
\label{fig:adscat}}
\end{figure}
%%%%%%%%%%%%%%%%%%%%%%%%%%%%%%%%%%%%%%%%%%%%%%%%%%%%%%%%%%%%%%%%%%%%%%%%%%%%%%%%%%%%%%%%%%%%%%%

If $a>0$,
the truncated 1PI effective action $\Gamma_{1+2} [A]$
gives nontrivial predictions
for atom-dimer scattering
that can be obtained by using the LSZ formalism.
These predictions can be obtained
more directly
from the equivalent truncated effective action
$\Gamma_{1+2} [A,D]$
in Eq.~(\ref{GammaAD}) with a classical dimer field.
Atom-dimer scattering proceeds through the Feynman
diagram in Fig.~\ref{fig:adscat}.
The $T$-matrix element
obtained using the Feynman rules
in Fig.~\ref{fig:rules-AD} is
\begin{equation}
{\cal T} (k, \theta) \approx
\frac {16 \pi a /m}
{1 + (5 + 4 \cos \theta) \ a^2 k^2 /4},
\label{Tmatrix:Simplify}
\end{equation}
where $k$ is the momentum of the atom
or the dimer in the center-of-mass frame and
$\theta$ is the scattering angle.
The atom-dimer scattering length $a_{AD}$
can be obtained from the T-matrix element
at $k=0$. Using
${\cal T} (k, \theta) \to - 3 \pi a_{AD} /m$
as $k \to 0$,
we obtain
\begin{equation}
a_{AD} = - \frac{16}{3} a.
\label{aAD-1}
\end{equation}
The atom-dimer scattering length
has been calculated previously
in Ref.~\cite{RDSS}
using an approximation equivalent to
our truncated 1PI approximation.
Their result $32 a /3$ differs from ours
by a factor of $-2$.
Part of the discrepancy seems to come from
a factor of 2 error
in the expression for the $T$-matrix element $T_{am}$
associated with the diagram in Fig.~2(a) of Ref.~\cite{RDSS},
which is
the same as our Fig.~\ref{fig:adscat}.

Using the $T$-matrix element
in Eq.~(\ref{Tmatrix:Simplify}),
we can obtain the total elastic atom-dimer
cross section:
\begin{equation}
\sigma_{\rm AD} (E) =
\frac {1024 \pi a^2 /9}
{(1+ 3 m a^2 E) (1+ m a^2 E /3)},
\label{sigma:total}
\end{equation}
where $E = 3 k^2 / 4 m $ is the collision energy.
By applying the partial wave decomposition to the
$T$-matrix element in Eq.~(\ref{Tmatrix:Simplify}),
we can isolate the $S$-wave contribution to
this cross section:
%----------------------
\begin{equation}
\sigma_{\rm AD}^{(L=0)} (E) =
16 \ \pi a^2
\left[ \frac{1}{ma^2 E}
    \log \left( \frac{1 + 3 ma^2 E} {1 + ma^2 E /3} \right)
\right] ^2 .
\label{sigmaAD-E-Swave}
\end{equation}
%----------------------
%
At the dimer-breakup threshold
$E = E_D =1 /(m a^2)$,
the $S$-wave cross section is
%----------------------
\begin{equation}
\sigma_{\rm AD}^{(L=0)} (E_D)
= \left( 16 \ \log ^2 3 \right) \pi a^2
\approx 19.31 \ \pi a^2.
\label{sigmaAD-ED-Swave}
\end{equation}
%----------------------
The result in Eq.~(\ref{sigma:total})
for the total cross section
at this energy is
$(64/3) \pi a^2$.
Thus the truncated 1PI effective action
predicts that $S$-wave scattering should account for
$90.5 \%$
of the cross section at the dimer-breakup threshold.

The results from the truncated 1PI effective action
can be compared with those from
exact solutions of the 3-body problem.
The exact results that are known are summarized
in Ref.~\cite{Braaten:2004rn}.
The exact result
for the atom-dimer scattering length is
%------------------
\begin{equation}
a_{AD} = \big(
1.46 - 2.15
\tan \left[ s_0 \ln (a \kappa_*) + 1.16 \right]
\big) \ a,
\label{aAD}
\end{equation}
%------------------
where $ s_0=1.00624$
and $\kappa_*$ is the Efimov parameter.
The functional form was first deduced
by Efimov~\cite{Efimov:1979}.
The numerical coefficients were first calculated
by Simenog and Sitnichenko~\cite{Sim81}
and by Bedaque, Hammer, and van~Kolck~\cite{BHK99,BHK99b}.
The ratio $a_{AD} /a$ from Eq.~(\ref{aAD})
can have any value
between $- \infty$ and $+ \infty$ depending on
the value of the product $a \kappa_*$.
Thus the result for $a_{AD}$ in Eq.~(\ref{aAD-1})
from the truncated 1PI effective action
can have an arbitrarily large error.

Another exact result for atom-dimer scattering is
the $S$-wave phase shift
at the dimer-breakup threshold,
which was derived by
Macek, Ovchinnikov, and Gasaneo~\cite{MOG05}.
The $S$-wave contribution to the cross section
at the dimer-breakup threshold is
%----------------------
\begin{equation}
\sigma_{\rm AD} ^{(L=0)} (E_D)
= 3 \pi a^2  \, \sin ^2
\left[ s_0 \ln (a \kappa_*) + 2.91 \right] .
\label{sigmaAD-ED-exact}
\end{equation}
%----------------------
The exact cross section can have any value
between $0$ and $3 \pi a^2$ depending on
the value of the product $a \kappa_*$.
The $S$-wave cross section from
the truncated 1PI effective action
in Eq.~(\ref{sigmaAD-ED-Swave})
is about $6$ times larger than
the maximum value of the exact cross section.
We see again that
the truncated 1PI effective action
can give results
for low-energy 3-body processes
with large errors.

\section{Static Homogeneous States}
\label{ch4}

In this section and the next one,
we use the truncated 1PI approximation
as a model for
a Bose gas of atoms
with large scattering length $a$ and
total atom number density $n$
at zero temperature.
For $a > 0$, the results illustrate
the equivalence of the truncated 1PI effective actions
$\Gamma_{1+2}[A; \mu]$
and $\Gamma_{1+2}[A,D; \mu]$.
In this section, we study
the static homogeneous states
predicted by this model.

\subsection{Atom condensate}

We can study the static homogeneous states
of the system using
the truncated 1PI effective action
$\Gamma_{1+2} [A; \mu]$ in Eq.~(\ref{GammaA}).
In a static homogeneous state,
the classical atom field
is a complex constant:
$A({\bf r},t) = \bar A$.
We will refer to $\bar A$
as the {\it atom mean field}.
A state with $\bar A \not= 0$ contains
a Bose-Einstein condensate of atoms.
In such a state, a macroscopic fraction
$|A|^2 /n$
of the atoms have
the same constant wavefunction
whose phase is that of $\bar A$.

The atom mean field $\bar A$ must satisfy
Eq.~(\ref{varyA}), which reduces to
\begin{equation}
\mu \bar A =
\frac {4 \pi a}
{1 - r}
|\bar A|^2 \bar A .
\label{varyA:bar}
\end{equation}
%
%[dz, eb]
We have introduced
a dimensionless chemical potential
variable $r$ defined by
\begin{equation}
r = a (-2 \mu - i \epsilon)^{1/2}.
\label{r}
\end{equation}
%
%[eb,dz]
We assume that $\mu$ is negative
so $r$ is real-valued
and has the same sign as $a$.
The expression for the chemical potential in terms of $r$ is
\begin{equation}
\mu = - r^2 / (2 a^2).
\label{mu-r}
\end{equation}
%
%[eb,dz]
%
If Eq.~(\ref{varyA:bar})
has a nonzero solution $\bar A$,
the atom mean field must satisfy
\begin{equation}
|\bar A|^2 = {r^2 (r - 1) \over 8 \pi a^3}.
\label{Abar2}
\end{equation}
%
%[eb,dz]
Since the left side of Eq.~(\ref{Abar2}) is
real and positive,
the right side must also be real and positive.
This requires $r<1$ if $a<0$ and $r>1$ if $a>0$.
However, if $a<0$, Eq.~(\ref{r}) with $\mu <0$ sets
a stronger constraint $r<0$.
Thus a static homogeneous state with
non-zero $\bar A$ is possible only if
$r<0$ or $r>1$.

The number density $n$ and the free energy density $\cal F$
in the static homogeneous state are given by
setting $A({\bf r},t) = \bar A$
in Eqs.~(\ref{n-A}) and (\ref{F-A}):
\begin{subequations}
\begin{eqnarray}
n &=& | \bar A |^2
+ {2 \pi a^3 \over  r (r-1)^2} | \bar A |^4 ,
\label{nAA}
\\
{\cal F} &=&
{r^2 \over 2 a^2}| \bar A |^2 - {2 \pi a \over r-1} | \bar A |^4 .
\label{FAA}
\end{eqnarray}
\label{nFAA}
\end{subequations}
%
%[eb,dz]
Using Eq.~(\ref{Abar2}) to eliminate $| \bar A |^2 $,
we obtain parametric expressions for $n$ and $\cal F$
as functions of $r$:
\begin{subequations}
\begin{eqnarray}
n &=& {r^2 ( 5r - 4)  \over 32 \pi a^3} ,
\label{n-r}
\\
{\cal F} &=& {r^4(r-1) \over 32 \pi a^5} .
\label{F-r}
\end{eqnarray}
\label{nF-r}
\end{subequations}
%
%[eb,dz]
If $a<0$, the number density $n$ increases from $0$ to $\infty$
as $r$ decreases from $0$ to $-\infty$.
If $a > 0$, the condition $r>1$
implies that $n$ is greater than the critical number density
\begin{equation}
n_c = 1 / (32 \pi a^3) .
\label{nc}
\end{equation}
%
%[eb,dz]
%
Equivalently it implies that
$a$ is greater than the critical
scattering length
\begin{equation}
a_c = (32 \pi n)^{-1/3} .
\label{ac}
\end{equation}
The number density in Eq.~(\ref{n-r})
increases from $n_c$ to $\infty$
as $r$ increases from 1 to $\infty$.

The two terms in $n$ and in ${\cal F}$
in Eqs.~(\ref{nFAA})
can be interpreted as
condensate and non-condensate contributions, respectively.
Dividing Eq.~(\ref{Abar2}) by Eq.~(\ref{n-r}),
we obtain the fraction of the atoms that are in
the atom condensate:
%------------------
\begin{eqnarray}
|\bar A|^2 /n =
\frac{4 (r-1)} {5 r - 4}.
\label{frac-A}
\end{eqnarray}
%------------------
%
If $a<0$, this fraction increases
from $0$ to $4/5$ as $n$ increases
from $0$ to $\infty$.
If $a>0$, this fraction increases
from $0$ to $4/5$ as $n$ increases
from $n_c$ to $\infty$.

%%%%%%%%%%%%%%%%%%%%%%%%%%%%%%%%%%%%%%%%%%%%%%%%%%%%%%%%%%%%%%%%%%%%%%%%%%%%%%%%%%%%%%%%%%%%%%%
\begin{figure}
\centerline{\includegraphics[width=10cm,angle=270] {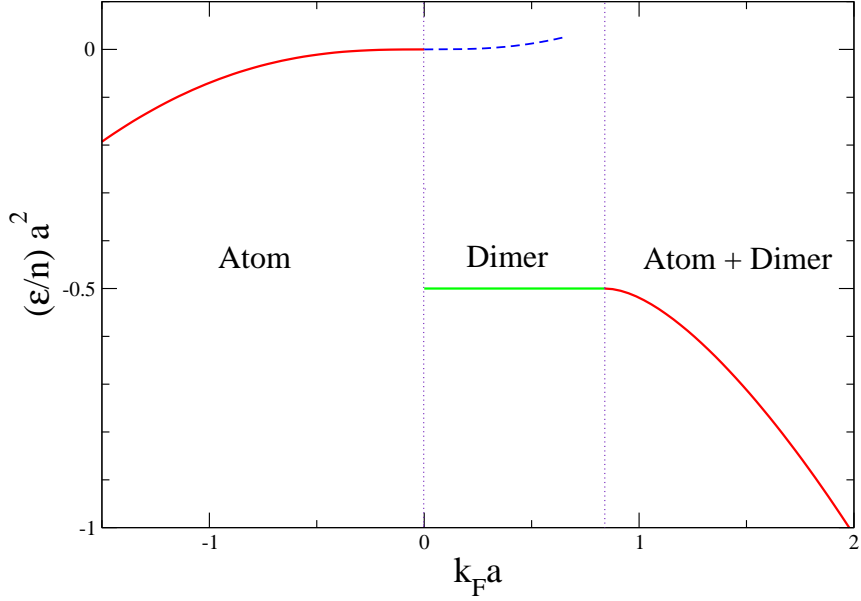} }
\caption{ Energy per particle ${\cal E} / n$
in units of $1/a^2$
as functions of $ k_{F} a $,
where $k_{F} = (6 \pi ^2 n)^{1/3}$.
The vertical dotted lines
separate the regions of $ k_{F} a $
in which the static homogeneous state includes
an atom condensate $( a < 0 )$,
a dimer condensate $( 0 < a  < a_c )$,
and a mixture of atom and dimer condensates
$( a  > a_c )$.
The dashed line represents an unstable
homogeneous state that includes
an atom condensate.
\label{fig:Phase-Weak} }
\end{figure}
%%%%%%%%%%%%%%%%%%%%%%%%%%%%%%%%%%%%%%%%%%%%%%%%%%%%%%%%%%%%%%%%%%%%%%%%%%%%%%%%%%%%%%%%%%%%%%%

The energy density
in the static homogeneous state is
${\cal E}={\cal F} + \mu n$,
%
%\begin{equation}
%{\cal E} = - {r^4(3r-2) \over 64 \pi a^5} .
%\label{E-r}
%\end{equation}
%
%[eb,dz]
where $\mu$, $n$, and ${\cal F}$
are given in
Eqs.~(\ref{mu-r}) and (\ref{nF-r}).
The average energy per atom is
\begin{equation}
{\cal E}/n = - {r^2(3r-2) \over 2(5r - 4)a^2} .
\label{En-r}
\end{equation}
%
%[eb,dz]
The energy per atom is shown as
a function of $a$
in Fig.~\ref{fig:Phase-Weak}.
In the high-density limit, a convenient energy
scale is the Fermi energy $\epsilon_F$ for an ideal gas of fermions
with a single spin component:
$\epsilon_F = \mbox{$\frac{1}{2}$} (6 \pi ^2 n)^{2/3}$.
%
%k_F = (6 \pi ^2 n)^ {1/3} ,
%\label{kF}
%[dz]
%
The energy per atom in Eqs.~(\ref{En-r})
has a well-defined unitary limit
as $a \to \pm \infty$ with $n$ fixed:
%------------------
\begin{eqnarray}
{\cal E} / n \longrightarrow
-\frac{3}{5} \left( \frac{16}{15 \pi} \right)^{2/3} \epsilon_F
\equiv \epsilon_{\infty} .
\label{unitaryE}
\end{eqnarray}
%------------------
%$-{3 \over 10} ({32 \over 5} \pi n)^{2/3}$.
%
Its numerical value is
$\epsilon_{\infty} \approx - 0.292 \ \epsilon_F .$
If $a < 0$, ${\cal E}/n$ decreases
from $2 \pi n a$ to $\epsilon_{\infty}$
as $n$ increases from $0$ to $\infty$.
The limiting value $2 \pi n a$ as $n \to 0$
is the analytic continuation to negative $a$
of the energy per particle of a weakly interacting
Bose gas.
If $a > 0$, ${\cal E}/n$ decreases
from $- 1/(2 a^2)$ to $\epsilon_{\infty}$
as $n$ increases from $n_c$ to $\infty$.
The limiting value $- 1/(2 a^2)$ at $n=n_c$
is the energy per atom for a pair of atoms
bound into a dimer with binding energy
$E_D = 1 /a^2$.

%\subsection{Unstable atom condensate}

If $a>0$ and $0 < n < n_c$,
there are unphysical solutions to
Eqs.~(\ref{varyA:bar}) and (\ref{nAA}).
The expression for $n$ in Eq.~(\ref{n-r})
is a cubic polynomial in $r$.
It therefore has three roots:
a real root in the range ${4\over5} < r < 1$
and a pair of complex roots.
For the real root, the expression
for $|\bar A|^2$ in Eq.~(\ref{Abar2})
has an unphysical negative value.
For the complex roots, the expression
for $|\bar A|^2$ in Eq.~(\ref{Abar2})
has an unphysical imaginary part.
If $n \ll n_c$, the complex roots
can be expanded in powers of $n a^3$:
\begin{eqnarray}
r &=& \pm i (8 \pi n a^3)^{1/2}
\left[ 1 \pm i \mbox{$5 \over 8$} (8 \pi n a^3)^{1/2}
+ \ldots \right].
\label{r-na3}
\end{eqnarray}
%
%[dz]
The thermodynamic quantities corresponding to this root
can also be expanded in powers of $n a^3$.
The expansion for the energy per atom is
\begin{eqnarray}
{\cal E}/n = 2 \pi n a \left[
1 \pm i (8 \pi n a^3)^{1/2}
+ \ldots \right].
\label{EoverN-na3}
\end{eqnarray}
%
%[dz]
%
The leading term in Eq.~(\ref{EoverN-na3})
is the well-known result for the
weakly-interacting Bose gas.
If $n a^3$ is small, the imaginary part is suppressed by
a factor of $(n a^3)^{1/2}$.
If $a$ is decreased through $0$
to a negative value,
one of the roots in Eq.~(\ref{r-na3})
becomes the negative value of $r$
that corresponds to the static homogeneous state
with mean-field $\bar A$ satisfying
Eq.~(\ref{Abar2}).
Thus the weakly-interacting Bose gas
can be interpreted as the analytic continuation
to $a>0$
of the static homogeneous state for $a<0$.
The real part of ${\cal E}/n$
as a function of $a$
is shown as a dashed line in
Fig.~\ref{fig:Phase-Weak}.

\subsection{Mixture of atom and dimer condensates}

If $a > 0$,
we can also study the static homogeneous states
of the system using
the truncated 1PI effective action $\Gamma_{1+2}[A,D; \mu]$
given in Eq.~(\ref{GammaAD}).
In a static homogeneous state,
the classical atom and dimer fields
must be complex constants:
$A({\bf r},t) = \bar A$, $D({\bf r},t) = \bar D$.
We will refer to $\bar A$ and $\bar D$
as the atom and dimer mean fields, respectively.
A state with $\bar A \neq 0$ and $\bar D \neq 0$
contains a mixture of Bose-Einstein condensates of
atoms and dimers.
In such a state, a macroscopic fraction
$|\bar A|^2 /n$
of the atoms have the same constant wavefunction
whose phase is that of $\bar A$.
A macroscopic fraction
$2 |\bar D|^2 /n$
of the atoms
are bound into dimers that have the same constant
center-of-mass wavefunction
whose phase is that of $\bar D$.

The atom and dimer mean fields $\bar A$ and $\bar D$
must satisfy Eqs.~(\ref{varyAD}), which reduce to
\begin{subequations}
\begin{eqnarray}
\mu \bar A
&=& { 16 \pi \over a g} \bar A^* \bar D
- \frac {4 \pi a} {1 + r}
|\bar A|^2 \bar A ,
\label{varyAD:1bar}
\\
2 \mu \bar D &=&
- {1 \over a^2} \bar D + \mbox{$1 \over 2$} g \bar A^2 ,
\label{varyAD:2bar}
\end{eqnarray}
\label{varyAD:bar}
\end{subequations}
%
%[eb,dz]
where $r$ is the dimensionless
chemical potential variable defined in Eq.~(\ref{r}).
Eq.~(\ref{varyAD:2bar}) can be solved for the mean field
$\bar D$ as a function of $\bar A$:
\begin{equation}
\bar D = - {g a^2 \over 2 (r^2-1)} \bar A^2 .
\label{Dbar}
\end{equation}
%
%[eb,dz]
Inserting this expression for $\bar D$ into
Eq.~(\ref{varyAD:1bar})
and solving for $|\bar A|^2$, we obtain
the same
expression as in Eq.~(\ref{Abar2}),
which was derived from the action $\Gamma_{1+2}[A; \mu]$.

The number density $n$ and the free energy density $\cal F$
in the static homogeneous state are given by
Eqs.~(\ref{n-AD}) and (\ref{F-AD}):
\begin{subequations}
\begin{eqnarray}
n &=& | \bar A |^2 + 2 | \bar D |^2
+ {2 \pi a^3 \over r (r+1)^2}  | \bar A |^4 ,
\label{nAD}
\\
{\cal F}  &=& {r^2 \over 2 a^2}  |\bar A|^2
+ {r^2 - 1 \over a^2} |\bar D |^2
+ g \, {\rm Re}( \bar D^* \bar A^2 )
- {2 \pi a \over r+1} | \bar A |^4 .
\label{FAD}
\end{eqnarray}
\label{nFAD}
\end{subequations}
%
%[dz,eb]
The first two terms in $n$ and in ${\cal F}$ can be interpreted
as atom condensate and dimer condensate contributions, respectively.
The remaining terms can be interpreted as
non-condensate contributions.
Using Eqs.~(\ref{Dbar}) and (\ref{Abar2}) to eliminate
the mean fields $\bar D$ and $\bar A$ in favor of $r$ and $a$,
we obtain
the same expressions for $n$ and $\cal F$
as in Eqs.~(\ref{nF-r}),
which were derived from the action $\Gamma_{1+2}[A; \mu]$.

Nonzero values of
$\bar A$ and $\bar D$
that satisfy the variational equations for
$\Gamma_{1+2}[A,D; \mu]$
correspond to a state
that contains a mixture of
an atom condensate
and a dimer condensate.
One might be tempted to interpret
a nonzero value of $\bar A$
that satisfies the variational equations for
$\Gamma_{1+2}[A; \mu]$
as the atom mean field in a state with
an atom condensate
but no dimer condensate.
In this case it would
seem quite remarkable that
the static homogeneous state with nonzero
$\bar A$ and $\bar D$ derived from $\Gamma_{1+2}[A,D; \mu]$
has exactly the same thermodynamic properties
as the static homogeneous state with the same value of
$\bar A$ derived from $\Gamma_{1+2}[A; \mu]$.
However the functionals
$\Gamma_{1+2}[A; \mu]$ and $\Gamma_{1+2}[A,D; \mu]$
are equivalent.
Thus the state with nonzero $\bar A$ and $\bar D$
derived from $\Gamma_{1+2}[A,D; \mu]$ is in fact
identical to the state
with the same value of $\bar A$
derived from $\Gamma_{1+2}[A; \mu]$.
Constructing the equivalent action
$\Gamma_{1+2}[A,D; \mu]$
and looking for solutions with
a nonzero value of $\bar D$
as well as a nonzero value of $\bar A$
does not reveal any new states of the system.
However it does allow
non-atom-condensate contributions
to thermodynamic quantities to be resolved into
dimer condensate contributions
and contributions that correspond to neither
atom nor dimer condensates.

Using Eqs.~(\ref{Dbar}), (\ref{Abar2}),
and (\ref{n-r}),
we can determine the fraction of the atoms
that are
in the dimer condensate:
%------------------
\begin{eqnarray}
2 |\bar D|^2 /n =
\frac{4 r^2} {(r+1)^2 (5 r - 4)}.
\label{fraction-D}
\end{eqnarray}
%------------------
%
This fraction decreases
from $1$ at $n = n_c$
to $0$ as $n \to \infty$.

\subsection{Dimer condensate}

If $a>0$, we have found that for number density $n$
less than the critical value $n_c$ defined in Eq.~(\ref{nc}),
there are no static homogeneous states with
$\bar A \neq 0$.
However, there is another possibility.
If the chemical potential has the value
\begin{equation}
\mu = - 1 / (2a^2),
\label{mu:pureD}
\end{equation}
%
%[eb,dz]
the variational Eqs.~(\ref{varyAD:bar})
can be satisfied
if $\bar A = 0$ and $\bar D \not= 0$.
This corresponds to a state
that has a dimer condensate but
no atom condensate.
Since $r=1$, the number density
and the free energy density in Eqs.~(\ref{nFAD})
reduce to
\begin{subequations}
\begin{eqnarray}
n &=&  2 |\bar D |^2,
\label{n:pureD}
\\
{\cal F} &=& 0.
\label{F:pureD}
\end{eqnarray}
\label{nF:pureD}
\end{subequations}
%
%[dz,eb]
The energy density is therefore
${\cal E} = \mu n$,
and the average energy per atom is
\begin{equation}
{\cal E} /n = - 1 / (2a^2).
\label{EoverN:pureD}
\end{equation}
%
%[dz,eb]
%
The energy per atom in this state
is shown as a horizontal line
in Fig.~\ref{fig:Phase-Weak}.
Since the number density of dimers is $n/2$,
the energy per dimer is $-1/a^2$.
Both $\mu$ and ${\cal E} /n$ have the same values
as a gas consisting of dimers
that have the binding energy
$E_D = 1/a^2$
but are otherwise noninteracting.
Thus we can interpret this state
as a pure dimer condensate.

\subsection{Discussion of phases}

The truncated 1PI approximation predicts
one static homogeneous state
for any value of the scattering length $a$.
%as shown in Fig.~\ref{fig:Phase-Weak}.
%
We can interpret the various static homogeneous states
as different superfluid phases
separated by phase transitions
at $a=0$ and at the critical value $a=a_c$
defined in Eq.~(\ref{ac}):
\begin{itemize}
\item{}
an atom superfluid phase with only an atom condensate for
$- \infty < a < 0$,
\item{}
a dimer superfluid phase with only a dimer condensate for
$0 < a < a_c$,
\item{}
an atom superfluid phase with a mixture of atom and dimer condensates for
$a_c < a < + \infty$.
\end{itemize}
The phase transitions are illustrated in
Fig.~\ref{fig:Phase-Weak},
which shows the average energy per atom
${\cal E} /n$
as a function of the dimensionless variable
$k_F a$, where $k_F = (6 \pi^2 n)^{1/3}$ is the Fermi wavenumber of
a system of fermions with
a single spin state
and number density $n$.
The energy per particle is given by
Eq.~(\ref{En-r})
for $- \infty < a < 0$,
by Eq.~(\ref{EoverN:pureD})
for $0 < a < a_c$,
and by Eq.~(\ref{En-r})
again for $a_c < a < + \infty$.
The energy per particle is discontinuous
at $a = 0$
and its second derivative is discontinuous
at $a = a_c$.

We proceed to determine the prediction
of the truncated 1PI effective action
for the orders of
the two phase transitions.
The free energy density $\cal F$ is given
by Eq.~(\ref{F-r})
for $- \infty < a < 0$,
by Eq.~(\ref{F:pureD})
for $0 < a < a_c$,
and by Eq.~(\ref{F-r})
again for $a_c < a < + \infty$.
The limiting behavior of $\cal F$
in the atom superfluid phase
as $a$ approaches the phase transition
at $a = 0$ is
%------------------
\begin{equation}
{\cal F} \longrightarrow - 2 \pi a n^2
\qquad {\rm as} \:
a \to 0^- .
\label{QFT-zero}
\end{equation}
%------------------
%
Since ${\cal F} = 0$
in the dimer superfluid phase,
${\cal F}$ is continuous at $a=0$
but its first derivative with respect to $a$
is discontinuous.
Thus the phase transition at $a=0$
is predicted to be first order.
The limiting behavior of $\cal F$
in the atom superfluid phase
as $a$ approaches the phase transition
at $a = a_c$ is
%------------------
\begin{equation}
{\cal F} \longrightarrow
\frac{96 \pi} {7} n^2 (a - a_c)
\qquad {\rm as} \:
a \to a_c ^+ .
\label{QFT-ac}
\end{equation}
%------------------
%
Since ${\cal F} = 0$
in the dimer superfluid phase,
${\cal F}$ is continuous at $a=a_c$
but its first derivative with respect to $a$
is discontinuous.
Thus the phase transition at $a=a_c$
is also predicted to be
first order.

%%%%%%%%%%%%%%%%%%%%%%%%%%%%%%%%%%%%%%%%%%%%%%%%%%%%%%%%%%%%%%%%%%%%%%%%%%%%%%%%%%%%%%%%%%%%%%%
\begin{figure}
\centerline{ \includegraphics[width=10cm,angle=270] {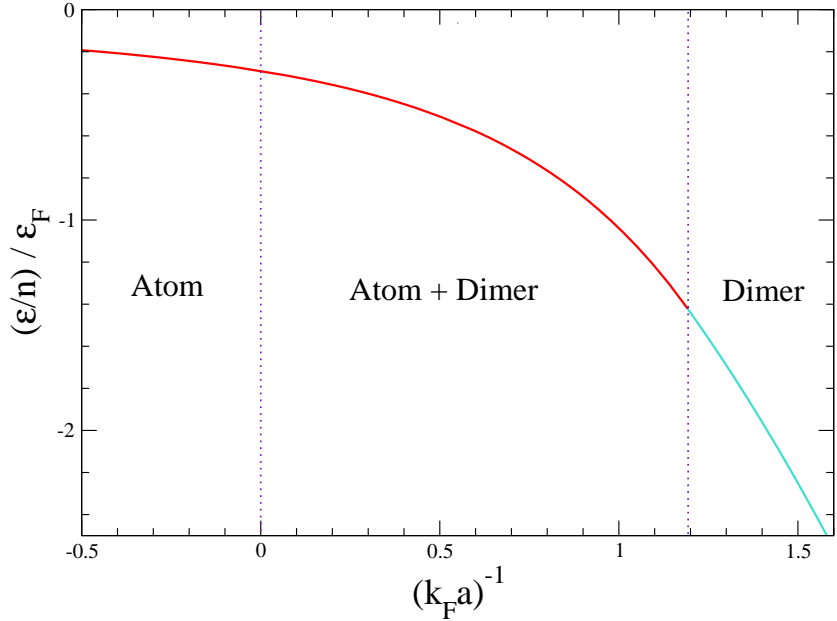} }
\caption{ Energy per particle ${\cal E} / n$
in units of the Fermi energy
$\epsilon_F = k_{F}^{2}/2$
as functions of $(k_{F} a)^{-1}$, where
$k_{F} = (6 \pi ^2 n)^{1/3}$.
The vertical dotted lines separate the
regions of $(k_{F} a)^{-1}$
in which the static homogeneous state includes
an atom condensate $( a^{-1} < 0 )$,
a mixture of atom and dimer condensates
$( 0 < a^{-1} < a_c^{-1} )$,
and a dimer condensate
$( a^{-1} > a_c^{-1} )$.
\label{fig:Phase-Strong} }
\end{figure}
%%%%%%%%%%%%%%%%%%%%%%%%%%%%%%%%%%%%%%%%%%%%%%%%%%%%%%%%%%%%%%%%%%%%%%%%%%%%%%%%%%%%%%%%%%%%%%%

%
The free energy density in Eq.~(\ref{F-r})
has the same limiting behavior as
$a \to + \infty$ and $a \to - \infty$,
so it has a well-behaved unitary limit:
%------------------
\begin{equation}
{\cal F} \longrightarrow
\frac{(32 \pi /5)^{2/3}} {5} n^{5/3}
\qquad {\rm as} \:
a \to \pm \infty.
\label{smooth-infty}
\end{equation}
%------------------
%
Not only is ${\cal F}$
a continuous function of $a^{-1}$
at $a^{-1} =0$,
but all derivatives of ${\cal F}$
with respect to $a^{-1}$
are also continuous.
Thus there is no phase transition at $a^{-1} = 0$
between the atom superfluid phase for $a^{-1} < 0$
and the atom superfluid phase for $a^{-1} > 0$.
The system in these two regions
differs in that
there is a dimer condensate
for $a^{-1} > 0$
and no dimer condensate
for $a^{-1} < 0$.
However the fraction of atoms
in the dimer condensate
goes to 0 in the unitary limit
in such a way
that all thermodynamic variables
remain smooth.
The smooth behavior of
the energy per atom as a function
of $a^{-1}$ at $a^{-1} = 0$
is illustrated in Fig.~\ref{fig:Phase-Strong}.

The smooth behavior of the thermodynamic variables
at $a^{-1} =0$
is consistent with the universality hypothesis
proposed by Ho~\cite{Ho-2004-92},
which asserts that the thermodynamic properties of
the system in the unitary limit $a = \pm \infty$
are universal functions of the number density $n$
and the temperature $T$
and don't depend on any interaction parameters.
This hypothesis requires that a thermodynamic variable
at $T = 0$ and $a = \pm \infty$
have the form $A \ n^p$, where $A$ is a constant
and $p$ is the power required
by dimensional analysis.
This hypothesis is well established
for systems consisting of
fermions with two spin states.
Ho proposed that it might also
apply to bosonic systems.

If a many-body system consisting of
identical bosons with large scattering length $a$
has number density $n$ and temperature $T$
such that $n^{-1/3}$ and $(m T / \hbar^2)^{-1/2}$
are large compared to the range,
its behavior should
be determined by $a$ and
the Efimov parameter $\kappa_*$
and it should be governed by
the discrete scaling symmetry.
Efimov physics requires a modification of
Ho's universality hypothesis~\cite{Ho-2004-92}
to allow for log-periodic dependence
on the Efimov parameter $\kappa_*$.
For example, a thermodynamic variable
at zero temperature in the unitary limit $a = \pm \infty$ must
have the form $n^p \ f(s_0 \log (n / \kappa_* ^3))$,
where $p$ is the power required by dimensional analysis
and $f(\theta)$ is a dimensionless function with period $2 \pi$.
Our truncated 1PI approximation
does not give this behavior, because irreducible 3-body amplitudes
have been neglected.

\subsection{Comparison with previous work}

Several groups have studied
the static homogeneous states for the many-body system
of bosonic atoms
near a Feshbach resonance
~\cite{RPW,RDSS,LL,BM}.
The starting point
for all these analyses could be taken to be
the quantum field theory with
an atom field $\hat \psi_A ({\bf r},t)$
and a molecule field $\hat \psi_M ({\bf r},t)$
defined by
the Hamiltonian density
in Eq.~(\ref{Res-AD}).
They considered the phase diagram of the system
as a function of the atom number density $n$,
the temperature $T$, and a detuning variable
that determines the scattering length $a$
of the atoms.
When the scattering length is large
compared to all other length scales,
the low energy behavior of the system
should be the same as in
the zero-range model.
We can compare our results from
the truncated 1PI effective action
at zero temperature with their results
for large scattering length.

We first consider the qualitative behavior
of the phase diagram at $T = 0$.
Radzihovsky, Park and Weichman (RPW) and
Romans, Duine, Sachdev and Stoof (RDSS)
independently pointed out that this system
has a quantum phase transition
between an atom superfluid phase
and a molecular superfluid phase~\cite{RPW,RDSS}.
%
%Basu and Mueller (BM) studied the effects of
%various instabilities on the phase transition~\cite{BM}.
%
The quantum phase transition is between
states with different symmetries.
The fundamental theory
defined by the Hamiltonian density
in Eq.~(\ref{Res-AD})
has a $U(1)$ symmetry
corresponding to phase transformations
of the atom and molecule quantum fields:
$\hat \psi_A \to e^{i \theta} \hat \psi_A$,
$\hat \psi_M \to e^{2 i \theta} \hat \psi_M$.
In the atom superfluid phase,
both the atom and molecule fields have
nonzero expectation values
$\langle \hat \psi_A \rangle$ and
$\langle \hat \psi_M \rangle$,
so the $U(1)$ symmetry is completely broken
to the trivial subgroup $\{ +1 \}$.
In the molecular superfluid phase,
$\langle \hat \psi_M \rangle$ is nonzero
but $\langle \hat \psi_A \rangle = 0$,
so the $U(1)$ symmetry is broken
to the subgroup $Z_2 = \{ +1, -1 \}$.
Thus the order parameter for the phase transition
is an element of the
discrete group $Z_2$~\cite{RPW,RDSS}.
For $a > 0$,
the classical atom and dimer fields
$A$ and $D$
in the truncated 1PI effective action
$\Gamma_{1+2}[A,D; \mu]$
play the same role as the
expectation values
$\langle \hat \psi_A \rangle$ and
$\langle \hat \psi_M \rangle$
in Refs.~\cite{RPW,RDSS}.
For $a < 0$, we have no analog of
$\langle \hat \psi_M \rangle$.
However
the classical atom field $A$
in the truncated 1PI effective action
$\Gamma_{1+2}[A; \mu]$
plays the same role
as the expectation value
$\langle \hat \psi_A \rangle$
in Refs.~\cite{RPW,RDSS}.

The order of the quantum phase transition
between the atom superfluid phase
and the molecular superfluid phase
has been discussed in Refs.~\cite{RPW,RDSS,LL}.
Because the two phases are distinguished by
a $Z_2$ order parameter, one might expect
the phase transition to be in the
universality class of the $4$-dimensional
Ising model, which has a second order transition.
However,
the coupling of the $Z_2$ order parameter to
the Goldstone mode associated with the
molecule condensate phase
modifies the critical behavior
and makes
the phase transition weakly first order.
The mean-field approximation
predicts that the phase transition
is second order.
Lee and Lee used the renormalization group
to show that the phase transition is
first order~\cite{LL}.
The truncated 1PI approximation
correctly predicts that the phase transition
is first order.

The analysis of the static homogeneous states
by RPW~\cite{RPW} was carried out
using the mean-field approximation,
in which the quantum field operators
in Eq.~(\ref{Res-AD})
are replaced by classical fields
$\psi_A ({\bf r},t)$
and $\psi_M ({\bf r},t)$
and the bare parameters,
which depend on the ultraviolet cutoff $\Lambda$,
are replaced by mean-field parameters.
The analysis of the static homogeneous states
by RDSS~\cite{RDSS} included some effects that
go beyond the mean-field approximation.
The grand Hamiltonian density
in the mean-field approximation is
\begin{eqnarray}
{\cal H} - \mu n &=& \psi_A ^*
\left(
- \mbox{$\frac{1}{2}$} \nabla ^2 - \mu
\right)
\psi_A
\, + \,
\psi_M ^*
\left(
- \mbox{$\frac{1}{4}$} \nabla ^2 - 2 \mu + \nu_M
\right)
\psi_M
\nonumber
\\
& + &
\mbox{$\frac{1}{4}$}
\lambda_{AA} \left(
\psi_A ^* \psi_A
\right)^2
\, + \,
\mbox{$\frac{1}{2}$}
g_{AM} \left(
\psi_A ^{*2} \psi_M +
\psi_M ^* \psi_A ^2
\right)
\nonumber
\\
& + &
\lambda_{AM}
\psi_M ^* \psi_A ^*
\psi_A \psi_M
\, + \,
\mbox{$\frac{1}{4}$}
\lambda_{MM} \left(
\psi_M ^* \psi_M
\right)^2 .
\label{H-AM}
\end{eqnarray}
The mean-field predictions
for the scattering observables
are sums of tree diagrams
whose vertices are given by the
interaction terms
in Eq.~(\ref{H-AM}).
For example, the $T$-matrix element
for atom-atom scattering is given by
the sum of the two diagrams in
Fig.~\ref{fig:ADA-scat},
where the vertices in the first diagram
are $-i g_{AM}$
and the vertex in the second diagram
is $-i \lambda_{AA}$.
The resulting expression for
the scattering length is
\begin{equation}
a = \frac{1} {8 \pi} \left(
\lambda_{AA} - \frac{g_{AM}^2} {\nu_M}
\right) .
\label{a-res}
\end{equation}
The resonance where the scattering length $a$
diverges is reached
by tuning the parameter
$\nu_M$ to 0.

We first discuss the values of the mean-field parameters
in Ref.~\cite{RPW}.
RPW stated that in the dilute gas limit,
$\lambda_{AA}$, $\lambda_{AM}$, $\lambda_{MM}$
are proportional to the scattering lengths
for atom-atom, atom-molecule,
and molecule-molecule scattering, respectively.
This presumably means that
the scattering length $a$ is given by
Eq.~(\ref{a-res})
with $g_{AM} = 0$.
This would be inconsistent with
the classical approximation.
RPW did not specify the parameter $g_{AM}$.
They also did not discuss the behavior of
the mean-field parameters near the resonance
where the scattering length $a$
is large.

The approximations of RDSS can also be expressed
in terms of an effective Hamiltonian with
the terms in Eq.~(\ref{H-AM})~\cite{RDSS}.
They discussed the behavior of
the mean-field parameters
near the resonance where
$a \to + \infty$.
For $a > 0$,
which corresponds to negative detuning,
the limiting behavior of the parameters is
\begin{subequations}
\begin{eqnarray}
\nu_{M} & \longrightarrow &
- 1/a^2,
\\
g_{AM} & \longrightarrow &
\sqrt{ 16 \pi /a },
\\
\lambda_{AA} & = &
8 \pi a_{bg},
\\
\lambda_{AM} & \longrightarrow &
32 \pi a,
\\
\lambda_{MM} & \longrightarrow &
16 \pi a .
\end{eqnarray}
\label{match-RDSS}
\end{subequations}
The limiting behavior of the detuning parameter
$\nu_M$ is the energy of the weakly-bound dimer.
The limiting value of $g_{AM}$
also comes from the exact solution
of the 2-body problem.
Note that all the coefficients in Eqs.~(\ref{match-RDSS})
scale as powers of $a$ except $\lambda_{AA}$.
RDSS took $\lambda_{AA}$ to be a constant,
so it gives
a non-resonant contribution $a_{bg}$
to the scattering length
in Eq.~(\ref{a-res}).
Note that if the limiting values of
$\nu_M$, $\lambda_{AA}$, and $g_{AM}$
in Eqs.~(\ref{match-RDSS})
are inserted into the right side of
Eq.~(\ref{a-res}),
the limiting value is $2 a$
instead of $a$.
RDSS took
$\lambda_{AM}$ and $\lambda_{MM}$
to be proportional to the scattering lengths
for atom-molecule and molecule-molecule scattering.
They recognized that
in the limit $a \to \infty$,
these scattering lengths must scale
linearly with
the large scattering length $a$.
RDSS did not recognize that
Efimov physics requires
the coefficients of $a$
in these scattering lengths to
be log-periodic functions of $a$
of the form $f(s_0 \log (a \kappa_*))$,
where $\kappa_*$ is the Efimov parameter and
$f(\theta)$ is a function with period $2 \pi$.
They calculated the atom-dimer and dimer-dimer
scattering lengths
using an approximation in which 2-body quantum effects
are treated exactly,
the atom-dimer coupling constant $g_{AM}$
is treated as a perturbation,
and other irreducible 3-body and 4-body quantum effects
are neglected.
In the case of the atom-dimer scattering length
$a_{AD}$,
this approximation is equivalent to
our truncated 1PI approximation.
Their result for $a_{AD}$
differs from ours in Eq.~(\ref{aAD-1})
by a factor of $-2$.
Both results for $a_{AD}$ differ from the exact result
in Eq.~(\ref{aAD})
in which the coefficient of $a$
is a log-periodic function of
$a \kappa_*$.
The approximation of RDSS
for the dimer-dimer scattering length is
$a_{DD} = 4 a$.
The exact result for $a_{DD}$
has not yet been calculated.

Our truncated 1PI approximation gives
predictions for the parameters
in the mean-field Hamiltonian
in Eq.~(\ref{H-AM})
in the limit of large scattering length.
The classical atom field $A({\bf r},t)$
in the 1PI effective action
can be naturally identified with
the atom mean field $\psi_A({\bf r},t)$.
If $a > 0$, we can also identify
the classical dimer field
$D({\bf r},t)$
with the molecular mean field
$\psi_M({\bf r},t)$.
Inserting constant mean fields
$\bar A$ and $\bar D$
into the truncated 1PI effective action
in Eq.~(\ref{GammaAD}),
it reduces to
\begin{eqnarray}
\Gamma_{1+2} [{\bar A},{\bar D}; \mu]
&=& \int dt \int d^3r \left\{
\mu {\bar A}^* {\bar A}
+ \left( 2 \mu + 1/a^2 \right) {\bar D}^* {\bar D}
\right.
\nonumber
\\
&& \hspace{2 cm} \left.
- \sqrt{4 \pi /a}
    \left({\bar A}^{*2} {\bar D}
    + {\bar D}^* {\bar A}^2 \right)
+ \frac{2 \pi a} { 1 + a \sqrt{-2 \mu} }
    \left( {\bar A}^* {\bar A} \right)^2
\right\} .
\label{Gamma-barAD}
\end{eqnarray}
We can identify the integrand
with the negative of the Hamiltonian density
in Eq.~(\ref{H-AM})
for constant values of the mean fields
$\psi_A$ and $\psi_M$.
The resulting predictions for
the mean-field parameters associated with 2-atom terms are
\begin{subequations}
\begin{eqnarray}
\nu_M & = & - 1/ a^2,
\\
g_{AM} & = & \sqrt{16 \pi /a},
\\
\lambda_{AA} & = &
\frac{- 8 \pi a} { 1 + a \sqrt{-2 \mu} }.
\end{eqnarray}
\label{param-match}
\end{subequations}
The values of $\nu_M$ and $g_{AM}$
agree with the limiting values of
RDSS near the Feshbach resonance.
Note that the value of $\lambda_{AA}$
depends on the chemical potential,
which depends on the number density.
If we take
the dilute limit in which $\mu \to 0$
and insert the parameters
in Eq.~(\ref{param-match})
into the right side of Eq.~(\ref{a-res}),
we get the correct
scattering length $a$.
Note that the scattering length is obtained
as the sum of two terms:
$a = (-a) + 2 a$.

A comparison of
Eqs.~(\ref{H-AM}) and (\ref{Gamma-barAD})
shows that the truncated 1PI approximation predicts
$\lambda_{AM} = 0$ and
$\lambda_{MM} = 0$.
These predictions
are artifacts of our truncation
of the 1PI effective action
after the 2-atom terms.
If the 1PI effective actions were truncated after
the 3-atom terms, there would be an additional term
$\lambda_{AM} {\bar D}^* {\bar A}^* {\bar A} {\bar D}$
in Eq.~(\ref{Gamma-barAD}).
If it were truncated after the 4-atom terms,
there would also be a term
$\mbox{$\frac{1}{4}$} \lambda_{MM} ( {\bar D}^* {\bar D} )^2$.
Using the exact result for
the atom-dimer scattering length
$a_{AD}$ in Eq.~(\ref{aAD}),
we can deduce the value of $\lambda_{AM}$.
The $T$-matrix element $i {\cal T}$
for atom-dimer scattering is the sum of
tree diagrams whose vertices are
1PI amplitudes.
Thus it is the sum of the diagram
in Fig.~\ref{fig:adscat}
whose contribution to the $T$-matrix element
is given in Eq.~(\ref{Tmatrix:Simplify})
and a diagram with a 1PI atom-dimer scattering amplitude.
For zero collision energy,
the 1PI atom-dimer scattering diagram
reduces to $-i \lambda_{AM}$
and the $T$-matrix element reduces to
${\cal T} = - 3 \pi a_{AD}$.
Thus the atom-dimer scattering length is
%------------------
\begin{equation}
a_{AD} = \frac{\lambda_{AM}}{3 \pi}
- \frac{16}{3} a .
\label{aAD-2}
\end{equation}
%------------------
Comparing with the exact result for
$a_{AD}$ in Eq.~(\ref{aAD}),
we obtain the exact result for $\lambda_{AM}$:
%------------------
\begin{equation}
\lambda_{AM} = 3 \pi \big(
6.79 - 2.15
\tan \left[ s_0 \ln (a \kappa_*) + 1.16 \right]
\big) \ a,
\label{lamAM-exact}
\end{equation}
%------------------
where $ s_0=1.00624$
and $\kappa_*$ is the Efimov parameter.

The location of the quantum phase transition
in the mean-field approximation
was determined in RPW and RDSS~\cite{RPW,RDSS}.
The critical number density $n_c$
at the quantum phase transition
satisfies%
\footnote
{In RPW~\cite{RPW}, the $g_{AM}$ term
in this equation has the wrong sign.}
%------------------
\begin{equation}
\nu_M = - g_{AM} \sqrt{2 n_c}
+ \left( \lambda_{AM}
- \mbox{$\frac{1}{4}$} \lambda_{MM}
\right) n_c .
\label{nc-eq}
\end{equation}
%------------------
If the dependence of the mean-field parameters
on the detuning parameter,
or equivalently,
on the scattering length $a$,
is known,
this condition can be translated into
a critical value $a_c$
of the scattering length.
To compare the condition in Eq.~(\ref{nc-eq})
with the results
from the truncated 1PI effective action,
we must set
$\lambda_{AM} = \lambda_{MM} = 0$.
Using the values of $\nu_M$ and $g_{AM}$
in Eq.~(\ref{param-match}),
we recover the expression
for the critical number density $n_c$
in Eq.~(\ref{nc}).

\section{ Quasiparticles }
\label{ch5}

In this section,
we use the truncated 1PI approximation
to study the quasiparticles
associated with small fluctuations around
the static homogeneous states.
There are two types of quasiparticles:
atom quasiparticles with
dispersion relation $\omega_A(k)$
and pair quasiparticles with
dispersion relation $\omega_P(k)$.
The two types of quasiparticles
can be distinguished by whether
the asymptotic behavior of their dispersion
relations as $k \to \infty$ is that for an atom
or a pair of atoms:
\begin{subequations}
\begin{eqnarray}
\omega_A(k) & \longrightarrow \mbox{$\frac{1}{2}$} k^2,
\label{omegaA}
\\
\omega_P(k) & \longrightarrow \mbox{$\frac{1}{4}$} k^2.
\label{omegaD}
\end{eqnarray}
\label{omegaAD}
\end{subequations}
%
%[dz, eb]
%
If the dispersion relation $\omega(k)$
for a quasiparticle has an imaginary part,
it indicates a dynamical instability of the system.

\subsection{Atom condensate}
\label{ch5.1}

To find the dispersion relation
$\omega (k)$ for quasiparticles associated with
fluctuations of $ A({\bf r},t)$
around the mean-field value $\bar A$, we look
for solutions to Eq.~(\ref{varyA})
of the form
\begin{equation}
A ({\bf r},t) = \bar A
+ \tilde A_+
e^{-i \omega t + i {\bf k} \cdot {\bf r}}
+ \tilde A_-
e^{+i \omega t - i {\bf k} \cdot {\bf r}} ,
\label{Atilde}
\end{equation}
%
%[eb,dz]
with $\tilde A_+$ and $\tilde A_-$ infinitesimal.
There are nontrivial solutions for
$\tilde A_+$ and $\tilde A_-$
only if the dispersion relations $\omega(k)$
satisfy
\begin{eqnarray}
0 &=& (a^2 \omega )^2 + r^2 (r-1) [ f_+ - f_-] a^2 \omega
- \mbox{$1 \over 2$} r^2 (r-1) [f_+ + f_-] (r^2 + a^2 k^2)
\nonumber
\\
&& - r^4 (r-1)^2 f_+ f_-
    - \mbox{$1 \over 4$} a^2 k^2 (2r^2 + a^2 k^2) ,
\label{freqA}
\end{eqnarray}
%
%[eb,dz]
where
\begin{equation}
f_\pm = {1 \over 1 - r (1 + {1 \over 4}a^2 k^2/r^2 \pm a^2
\omega/r^2 - i \epsilon)^{1/2} } \ .
\label{fpm}
\end{equation}
%
%[dz]
Since Eq.~(\ref{freqA}) is even in $\omega$,
the solutions come in pairs: if $\omega(k)$
is a solution, so is $- \omega(k)$.
The invariance of $\Gamma [A]$
under the $U(1)$ phase symmetry
$A \to e^{i\alpha} A$
guarantees that there must be a solution
that satisfies $\omega(k=0) = 0$.
This can be confirmed by observing that
$\omega = k =0$
is indeed a solution to Eq.~(\ref{freqA}).

Using algebraic manipulations, Eq.~(\ref{freqA})
can be transformed into a 6$^{\rm th}$ order
polynomial equation in $\omega^2$~\cite{zhang-thesis}.
The polynomial equation
is extremely complicated,
with 266 terms when fully expanded into
a polynomial in $\omega$, $k$ and $r$.
The explicit form is not very illuminating,
so we do not write it down here.
It is important that
Eq.~(\ref{freqA}) can be transformed into
a polynomial equation, because
it guarantees that we can find all the dispersion
relations that satisfy Eq.~(\ref{freqA}). A $6$th
order polynomial equation in $\omega ^2$ has
precisely $6$ roots in the complex $\omega ^2$ plane.
Given numerical values of $r$ and $k$,
the $6$ roots of the polynomial equation for $\omega^2$
can be easily found numerically.
One can then substitute the $12$ roots for $\omega$ into
Eq.~(\ref{freqA}) to see
which ones correspond to quasiparticle
dispersion relations $\omega (k)$.

We proceed to consider various limits
in which the quasiparticle dispersion relations
that satisfy Eq.~(\ref{freqA})
can be obtained in closed form.

\subsubsection { Non-interacting limit }
\label{section-noninteract}

\begin{figure}
\centerline{ \includegraphics[width=10cm, angle=270]{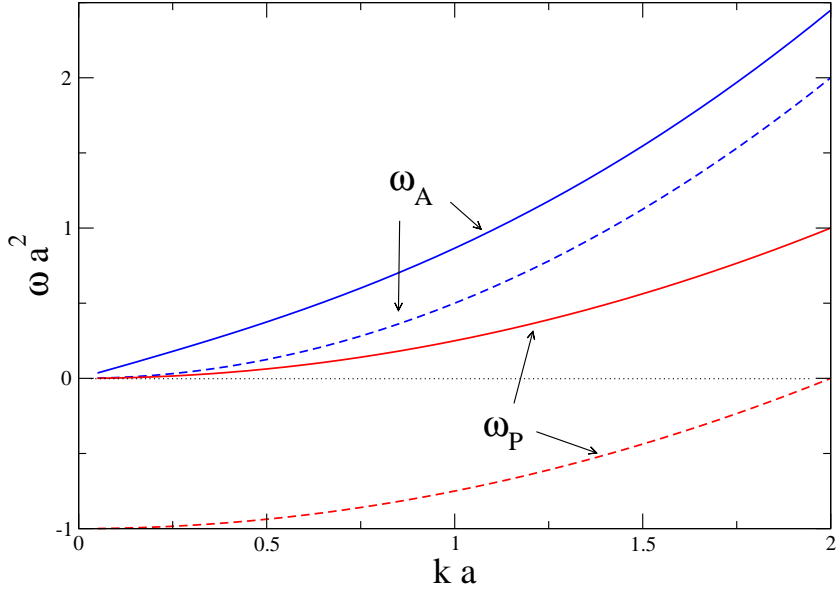} }
\caption{ Atom and pair dispersion relations
in the non-interacting limit ($r \to 0^-$, dashed line)
and the critical density ($r \to 1^+$, solid line).
\label{fig:r01}}
\end{figure}

In the limit $r \to 0^-$,
the quasiparticle dispersion relations
reduce to
\begin{subequations}
\begin{eqnarray}
\omega_A (k) &=& \mbox{$\frac{1}{2}$} k^2 ,
\label{omegaA:r0}
\\
\omega_P (k) &=& - 1/{a^2}
+ \mbox{$\frac{1}{4}$} k^2 .
\label{omegaD:r0}
\end{eqnarray}
\label{omegaAD:r0}
\end{subequations}
%
%[dz, eb]
These are the dispersion relation for an isolated atom
and for an isolated
dimer with binding energy $E_D = 1/a^2$, respectively.
The atom and pair dispersion relations in
Eqs.~(\ref{omegaAD:r0}) are
illustrated in Fig.~\ref{fig:r01} as dashed lines.
Since they are dispersion relations for
isolated atoms and dimers, we refer to
the limit $r \to 0^-$ as the
{\it non-interacting limit}.
Since $\omega_A (k)$ and $\omega_P (k)$ are real
for all $k$,
the homogeneous state is dynamically stable
in the non-interacting limit.

%%%%%%%%%%%%%%%%%%%%%%%%%%%%%%%%%%%%%%%%%%%%%%%%%%%%%%%%%%%%%%%%%%%%%%%%%%%%%%%%%%%%%%%%%%%%%%%
%%%%%%%%%%%%%%%%%%%%%%%%%%%%%%%%%%%%%%%%%%%%%%%%%%%%%%%%%%%%%%%%%%%%%%%%%%%%%%%%%%%%%%%%%%%%%%%

\subsubsection { Critical density $n_c$ }
\label{section-nc}

The limit $r \to 1^+$ corresponds to the critical density
$ n_c = 1/ (32 \pi a^3) $ defined in Eq.~(\ref{nc}).
In this limit, the quasiparticle dispersion relations
reduce to
\begin{subequations}
\begin{eqnarray}
\omega_A (k) &=& \mbox{$\frac{1}{2}$} k
\left( k^2 + 2/{a^2}
\right) ^{1/2} ,
\label{omegaA:r1}
\\
\omega_P (k) &=& \mbox{$\frac{1}{4}$} k^2 .
\label{omegaD:r1}
\end{eqnarray}
\label{omegaAD:r1}
\end{subequations}
%
%[dz, eb]
The atom dispersion relation is that
for the Bogoliubov mode in
a superfluid with coherence length $a/\sqrt{2}$.
The pair dispersion relation is that for
a dimer whose binding energy $E_D = 1/a^2$
is cancelled by its mean field energy.
The atom and pair dispersion relations in
Eqs.~(\ref{omegaAD:r1}) are
illustrated in Fig.~\ref{fig:r01} as solid lines.
Since $\omega_A (k)$ and $\omega_P (k)$
are real for all $k > 0$,
the homogeneous state is dynamically stable
at the critical density.

\subsubsection { Unitary limit }
\label{section-unitary}

The unitary limit is
$a \to \pm \infty$ with fixed number density $n$.
It corresponds to $r \to \pm \infty$
with $r/a = \sqrt{-2 \mu}$ fixed.
In this limit, it is
convenient to introduce
dimensionless scaling variables
for the frequency and wavenumber:
$\hat \omega = \omega a^2 / r^2$,
$\hat k = k a / r$.
After multiplying Eq.~(\ref{freqA}) by $r^{-4}$ and
taking the limit $r \to \pm \infty$, it reduces to
\begin{equation}
0 = {\hat \omega} ^2
+ ({\hat f}_{+} - {\hat f}_{-}) {\hat \omega}
- \mbox{$\frac{1}{2}$} ({\hat f}_{+} + {\hat f}_{-})
    (1 + {\hat k} ^2)
- {\hat f}_{+} {\hat f}_{-}
- \mbox{$\frac{1}{4}$} {\hat k}^2 (2 + {\hat k} ^2) ,
\label{scaling:freqA}
\end{equation}
where
\begin{equation}
{\hat f} _\pm = \frac{-1}
{
(1 + {1 \over 4} {\hat k}^2
\pm {\hat \omega} - i \epsilon)^{1/2}
}.
\label{scaling:fpm}
\end{equation}
%
%[dz]
Using algebraic manipulations,
Eq.~(\ref{scaling:freqA}) can be transformed into
a 6$^{\rm th}$ order polynomial equation
in $\hat \omega ^2$~\cite{zhang-thesis}.
The solutions must be inserted into
Eq.~(\ref{scaling:freqA})
to find which ones are quasiparticle
dispersion relations.
The atom and pair dispersion relations at large $\hat k$
have expansions in powers of $1/ \hat k$:
\begin{subequations}
\begin{eqnarray}
%k \to \infty:
\hat{\omega}_A(k) & = &
\frac{1}{2} \hat{k} ^2 + \frac{1}{2}
\pm i \frac{2}{\hat{k}}
+ \ldots ,
\label{freqA-unitary-kinfty}
\\
\hat{\omega}_P(k) & = &
\frac{1}{4} \hat{k} ^2 + 1
- \frac{16}{\hat{k} ^4}
+ \ldots
\label{freqD-unitary-kinfty}
\end{eqnarray}
\label{freq-unitary-kinfty}
\end{subequations}
The sign of the imaginary part of
$\hat{\omega}_A(k)$ is determined by
the choice of branch cut for the square root
in Eq.~(\ref{scaling:fpm}).
The terms of order $\hat{k} ^0$ in
Eqs.~(\ref{freq-unitary-kinfty})
imply that the atom and dimer have
mean-field energies
$| \mu |$ and $2 | \mu |$, respectively.
The limiting behaviors of
the dispersion relations
as $k \to 0$ are
\begin{subequations}
\begin{eqnarray}
\omega_A^2(k) & \longrightarrow &
(0.12 \pm i \ 0.45) \ k_F ^4 ,
\label{WA:res}
\\
\omega_P^2(k) & \longrightarrow &
 - \mbox{$\frac{1}{3}$}
(\mbox{$\frac{32}{5}$} \pi n)^{2/3} \, k^2,
\label{WD:res}
\end{eqnarray}
\label{WAD:res}
\end{subequations}
%
%[dz]
where $k_F = (6 \pi ^2 n)^{1/3}$
is the Fermi wavenumber.

%%%%%%%%%%%%%%%%%%%%%%%%%%%%%%%%%%%%%%%%%%%%%%%%%%%%%%%%%%%%%%%%%%%%%%%%%%%%%%%%%%%%%%%%%%%%%%%
\begin{figure}
\centerline{ \includegraphics[width=10cm,angle=270] {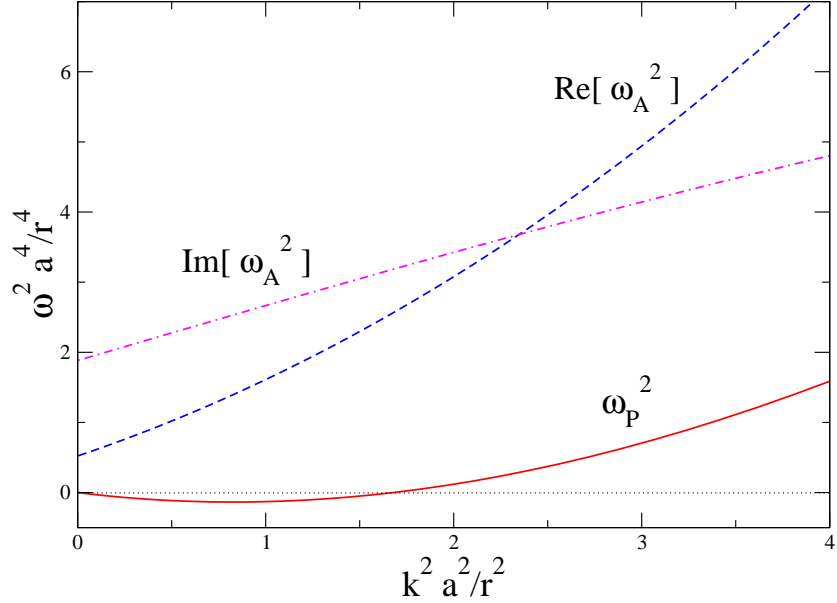} }
\caption{ The square of the pair dispersion relation (solid line)
and the real and imaginary parts of the square of the atom
dispersion relation (dashed line and dash-dotted line) in the
unitary limit $a \to \pm \infty$ with $n$ fixed.
\label{fig:unitary}}
\end{figure}
%%%%%%%%%%%%%%%%%%%%%%%%%%%%%%%%%%%%%%%%%%%%%%%%%%%%%%%%%%%%%%%%%%%%%%%%%%%%%%%%%%%%%%%%%%%%%%%

The atom and pair dispersion relations in the unitary limit
are illustrated in Fig.~\ref{fig:unitary} by plotting
${\rm Re} [ \hat{\omega}_A ^2 ]$,
${\rm Im} [ \hat{\omega}_A ^2 ]$,
and $\hat{\omega}_P ^2$
as functions of $\hat{k}^2$.
The atom dispersion relation
$\hat{\omega}_A$
is complex valued for all $\hat k$.
This indicates the system is unstable
under fluctuations
of any wavelength.
The pair dispersion relation
$\hat{\omega}_P$ remains real until
$k$ decreases
below the critical value
$k_{c} \approx 0.907 \, k_F$,
after which $\hat{\omega}_P$ becomes pure imaginary.

\subsubsection { General case }

Eq.~(\ref{freqA}) for the quasiparticle dispersion relations
cannot be solved analytically for general values of $r$
in the range $- \infty < r < 0$
and $1 < r < + \infty$.
However, analytic expression can be obtained
for limiting values of $k$.
At large $k$,
the atom and pair dispersion relations
have expansions in powers of $1/k$:
\begin{subequations}
\begin{eqnarray}
\omega_A(k) & = &
\mbox{$\frac{1}{2}$} k^2 + \frac{r^2}{2 a^2}
\pm i \frac{2 r^2 (r - 1)}{k a^3}
+ \cdots ,
\label{freqAkinfty}
\\
\omega_P(k) & = &
\mbox{$\frac{1}{4}$} k^2 + \frac{r^2 -1 }{a^2}
- \frac{8 r^2 (r - 1)}{k^2 a^4}
+ \cdots .
\label{freqDkinfty}
\end{eqnarray}
\label{freqADkinfty}
\end{subequations}
The terms of order $k^0$ imply that
the atom and pair quasiparticles
have mean-field energies
$\mbox{$\frac{1}{2}$} r^2 E_D$
and $r^2 E_D$, respectively.
Except at the limiting values
$r \to 0^-$ and $r \to 1^+$,
the atom dispersion relation
$\omega_A (k)$ is complex for all $k$.
Thus the system is unstable
under nonhomogeneous fluctuations
with any wavelength.
The pair dispersion relation $\omega_P (k)$
remains real until $k$ decreases below
a critical value $k_c$
that satisfies~\cite{zhang-thesis}
\begin{equation}
0 =
(k_c a)^6 + 4 (k_c a)^4 (r^2 - 1)
- 16 (k_c a)^2 r^2 (r - 1)
- 16 r^4 (r - 1)^2 ,
\label{kc}
\end{equation}
and then it becomes pure imaginary.
As $r$ increases from $1$ to $+ \infty$
and then increases from $- \infty$ to 0,
$k_c$ increases from $0$ to $0.907 \, k_F$
and then decreases to 0.

\subsection{Mixture of atom and dimer condensates}

To find the dispersion relations
$\omega (k)$ for quasiparticles associated with
fluctuations of $A({\bf r},t)$ and $D({\bf r},t)$
around the mean-field values $\bar A$ and $\bar D$,
we look for solutions
to Eqs.~(\ref{varyAD})
of the form
\begin{subequations}
\begin{eqnarray}
A ({\bf r},t) &=& \bar A
+ \tilde A_+
    e^{-i \omega t + i {\bf k} \cdot {\bf r}}
+ \tilde A_-
    e^{+i \omega t - i {\bf k} \cdot {\bf r}} ,
\\
D ({\bf r},t) &=& \bar D
+ \tilde D_+
    e^{-i \omega t + i {\bf k} \cdot {\bf r}}
+ \tilde D_-
    e^{+i \omega t - i {\bf k} \cdot {\bf r}} ,
\label{ADtilde}
\end{eqnarray}
\end{subequations}
%
%[dz]
with $\tilde A_+$, $\tilde A_-$,
$\tilde D_+$, and $\tilde D_-$ infinitesimal.
Surprisingly,
the condition for non-trivial solutions
can be reduced to Eq.~(\ref{freqA})~\cite{zhang-thesis}.
Thus the dispersion relations for
fluctuations of $A({\bf r}, t)$ and $D({\bf r}, t)$
predicted by Eq.~(\ref{varyAD})
are identical to the dispersion relations
for fluctuations of $A({\bf r}, t)$
predicted by Eq.~(\ref{varyA}).
This may seem quite remarkable at first,
but it is a consequence of
the equivalence of the effective actions
$\Gamma_{1+2}[A;\mu]$ and
$\Gamma_{1+2}[A,D;\mu]$.
Although these effective actions
give the same dispersion relations,
the coefficients of
$\tilde A_+$, $\tilde A_-$,
$\tilde D_+$, and $\tilde D_-$
obtained by solving
Eq.~(\ref{varyAD})
provide more detailed information
about the nature of the quasiparticles
than the coefficients of
$\tilde A_+$ and $\tilde A_-$
obtained by solving
Eq.~(\ref{varyA}).

\subsection{Pure dimer condensate}

If $a > 0$ and $0 < n < n_c$,
the static homogeneous state
is a dimer superfluid with mean fields
$\bar D \neq 0$ and $\bar A = 0$.
The equations for small oscillations of $A({\bf r},t)$
around $0$
and $D({\bf r},t)$ around $\bar D$ decouple.
The dispersion relations are
\begin{subequations}
\begin{eqnarray}
\omega_A (k) & = &
\left[
\left( 1/(2 a^2)
    + \mbox{$\frac{1}{2}$} k^2
\right)^2
- 8 \pi n/a
\right] ^{1/2} ,
\label{omega0A}
\\
\omega_P (k)& = &
\mbox{$\frac{1}{4}$} k^2 .
\label{omega0D}
\end{eqnarray}
\label{omega0AD}
\end{subequations}
%
%[eb,dz]
The pair dispersion relation in Eq.~(\ref{omega0D})
corresponds to a dimer whose binding energy
$E_D = 1/a^2$
is exactly canceled by its mean-field energy.
The atom dispersion relation in Eq.~(\ref{omega0A})
corresponds to a mean-field energy
$[1/(2 a^2)^2 - 8 \pi n/a ]^{1/2}$.
The mean-field energy vanishes at the critical density
$n_c = (32 \pi a^3)^{-1}$.

The dispersion relations in Eq.~(\ref{omega0AD})
are real-valued
for all $k$.
Thus if $a>0$,
the truncated 1PI approximation predicts that
the pure dimer condensate is
stable for $n < n_c$.
The existence of Efimov states suggests that if
3-body effects were taken into account,
the static homogeneous state containing
a dimer condensate but no atom condensate would also be unstable
for $n < n_c$.
Thus the stability of the dimer superfluid
is probably an artifact
of the truncated 1PI approximation.

\subsection{Comparison with previous work}

Several groups have studied
the quasiparticle dispersion relations
for the many-body system of bosonic atoms
near a Feshbach resonance using the mean-field
approximation~\cite{RPW,RDSS,BM}.
RPW and RDSS derived the quasiparticle dispersion relations
in the molecular superfluid phase~\cite{RPW,RDSS}.
Their dispersion relations reduce to
our results in Eqs.~(\ref{omega0AD})
if we set $\lambda_{AM} = \lambda_{MM} =0$ and
insert the values of the mean-field parameters
for the truncated 1PI approximation given
in Eqs.~(\ref{param-match}).
Basu and Mueller studied the quasiparticle dispersion relations
in the atom superfluid phase~\cite{BM}.
They showed that if the atom-molecule
and molecule-molecule coupling constants
$\lambda_{AM}$ and $\lambda_{MM}$
are set to zero,
there is a dispersion relation with the behavior
$\omega^2 \to c_s ^2 k^2$
as $k \to 0$
with $c_s ^2 < 0$.
Thus the atom superfluid phase is unstable with
respect to long-wavelength fluctuations.
Our result for the pair dispersion relation
$\omega_P (k)$ in the unitary limit
in Eq.~(\ref{WD:res})
is consistent with their result.
Our results go further by showing that
the atom dispersion relation $\omega_A (k)$
is complex for all $k$.
Thus the atom superfluid phase is unstable
with respect to fluctuations
of any wavelength.

\section { summary }

Previous work on the dimer superfluid phase
of the strongly-interacting Bose gas
has relied on the existence of a quantum field
whose expectation value can indicate the
presence of a dimer condensate.
The atom superfluid phase and the dimer superfluid phase
can both be described using the mean-field approximation.
However a dimer condensate is also possible in
models when there is no such quantum field.
In this case, the mean-field approximation cannot be used to
describe the dimer superfluid.
We have developed a method for describing the dimer
condensate in such case. It is based on the
1PI effective action $\Gamma[A]$, which is a functional
of classical atom field $A({\bf r},t)$
that encodes complete quantum information of the system.
The solutions to the variational equation for
$\Gamma [A]$
correspond to states that contains
an atom condensate only or a mixture of
an atom condensate and a dimer condensate.
The functional $\Gamma [A]$
also contains information about states
that contains a dimer condensate but no atom condensate,
but it is encoded in a more subtle way
through poles in functions of
differential operators.
In the case $a>0$, we constructed an equivalent
1PI effective action $\Gamma[A,D]$ that is also a functional of
a classical dimer field $D({\bf r},t)$.
Some of the solutions to the variational equation
for $\Gamma [A,D]$ correspond to the same states
as the solutions to the variational equation
for $\Gamma [A]$. These states contain an atom condensate only
or a mixture of an atom condensate and a dimer condensate.
However the variational equations for $\Gamma [A,D]$
has additional solutions that correspond to states
containing a dimer condensate but no atom condensate.

To illustrate our method for describing
the dimer condensate, we introduced an approximation
in which 1-atom and 2-atom terms are treated exactly
while irreducible $N$-atom terms with $N \ge 3$
are ignored.
This approximation can be defined by truncated 1PI effective actions
$\Gamma_{1+2}[A]$ and $\Gamma_{1+2}[A,D]$.
We determined these functionals exactly for the zero-range model
in which the only interaction parameter is
the large scattering length $a$.
We demonstrated the limitations of the truncated 1PI approximation
by using it to calculate the cross section for atom-dimer elastic scattering.
We compared the results with those from exact solutions of
the 3-body problem, which depend on the Efimov parameter $\kappa_*$.
The errors can be very large, so the truncated 1PI approximation
is not useful as a quantitative approximation.
However many results can be obtained analytically
in this approximation, so it may be useful as a benchmark
against which results from other methods can be compared.
We have found the truncated 1PI approximation to be particularly useful
for illustrating applications of the 1PI effective action.

We applied the truncated 1PI approximation to the
homogeneous strongly-interacting Bose gas
at zero temperature.
We first determined the static homogeneous states of the system.
The phase diagram is qualitatively consistent with
that derived in Refs.~\cite{RPW,RDSS}.
There is an atom superfluid phase and a dimer superfluid phase
separated by quantum phase transitions
at $a=0$ and $a_c = (32 \pi n)^{-1/3}$.
The unitary limit $a \to \pm \infty$ is consistent with
Ho's universality hypothesis~\cite{Ho-2004-92}.
There is a perfectly smooth transition from
an atom superfluid state containing a mixture of
an atom condensate and a dimer condensate
as $a \to + \infty$ to an atom superfluid
containing an atom condensate only as $a \to - \infty$.
We also determined the quasiparticles dispersion relations
associated with small fluctuations around
the homogeneous states.
In the atom superfluid state, the pair dispersion relation
$\omega_P (k)$
is pure imaginary for sufficiently small $k$,
indicating an instability with respect to
long-wavelength fluctuations.
The atom dispersion relation
$\omega_A (k)$
is complex for all $k$, indicating an instability
with respect to fluctuations
of any wavelength.
In the dimer superfluid state, the atom and pair dispersion relations
are real-valued for all $k$, indicating that the system is
dynamically stable.
We suggested that this stability
is an artifact of the truncated 1PI approximation.
In the case $a>0$, the properties of the atom superfluid phase
could be derived either from the variational equation
for $\Gamma_{1+2}[A]$ or from the variational equations
for $\Gamma_{1+2}[A,D]$.
They give exactly the same thermodynamic properties and
identical quasiparticle dispersion relations.
This illustrates the equivalence of
these two truncated 1PI effective actions.
Some advantages of $\Gamma_{1+2}[A,D]$ are that
it allows the non-condensate contributions to thermodynamic
properties to be separated into
a dimer-condensate contribution and a remainder
and it gives more detailed information about
the nature of the quasiparticles.
The properties of the dimer superfluid phase
can be derived easily from the variational equations
for $\Gamma_{1+2}[A,D]$.
That information must also be encoded in $\Gamma_{1+2}[A]$,
since these two functional are equivalent,
but is is in a much less accessible form.

Much of the previous work on the Bose gas
near a Feshbach resonance has been carried out
with the mean-field approximation using
the Hamiltonian density in Eq.~(\ref{H-AM}),
which depends on atomic and molecular mean fields
$\psi_A$ and $\psi_M$.
We used the truncated 1PI effective action
$\Gamma_{1+2}[A,D]$
to deduce the limiting values of some of the
mean-field parameters when the system is sufficiently
close enough to the Feshbach resonance.
Our results for $\nu_{M}$ and $g_{AM}$
agree with those of Ref.~\cite{RDSS}. Our result
for $\lambda_{AA}$
increases linearly with $a$, while $\lambda_{AA}$
was assumed to be constant in Ref.~\cite{RDSS}.
We also used results from exact solutions to the 3-body
problem to deduce the limiting value of $\lambda_{AM}$.
Our result for $\lambda_{AM}$
in Eq.~(\ref{lamAM-exact})
has log-periodic dependence on the Efimov parameter.
The simple result $\lambda_{AM} \to 32 \pi a$
in Ref.~\cite{RDSS}
was derived using an approximation equivalent to the truncated
1PI approximation.

The method we have used to describe
states containing a dimer condensate can be
generalized to states containing condensates of
Efimov trimers.
Information about a specific Efimov trimer is encoded
in the 1PI effective action $\Gamma[A]$
through the poles of functions of
differential operators that act on products
of three classical atom fields.
By introducing a classical trimer field $T({\bf r},t)$,
it should be possible to construct an equivalent
1PI effective action $\Gamma[A, T]$ or $\Gamma[A,D, T]$
in which these poles have been eliminated.
One could then use the variational equation for these
effective actions to study the trimer superfluid phase
of the strongly-interacting Bose gas.

\begin{acknowledgments}
%\Acknowledgement

We thank R.~Furnstahl and T.-L.~Ho for useful discussions.
This research was supported by DOE grants
DE-FG02-91ER4069 and DE-FG02-05ER15715.

\end{acknowledgments}

%%%%%%%%%%%% begin  bibliography  %%%%%%%%%%%%%%%%%%%%%%%%%

%%%%%%%%%%%% end  bibliography  %%%%%%%%%%%%%%%%%%%%%%%%%


\begin{thebibliography}{99}


%%%% BEC %%%%%

\bibitem{Anderson1995a}
M.~H.~Anderson {\it et al},
%``Observation of {B}ose-{E}instein Condensation in a Dilute Atomic Vapor''
Science {\bf 269}, 198-201 (1995).

\bibitem{Davis1995b}
K.~B.~Davis {\it et al},
%``{B}ose-{E}instein Condensation in a Gas of Sodium Atoms''
 Phys.\ Rev.\ Lett. {\bf 75}, 3969 (1995).

\bibitem{Bradley1995a}
C.~C.~Bradley {\it et al},
%``Evidence of {B}ose-{E}instein Condensation in an Atomic Gas with Attractive Interactions''
 Phys.\ Rev.\ Lett. {\bf 75}, 1687 (1995).


%%%% strongly-interacting Bose gas %%%%

\bibitem{Inouye98}
S.~Inouye, M.R.~Andrews, J.~Stenger, H.-J.~Miesner, D.M.~Stamper-Kurn,
        and W.~Ketterle,
%``Observation of Feshbach Resonances in a Bose-Einstein condensate,''
Nature {\bf 392}, 151 (1998).

\bibitem{Courteille98}
Ph.~Courteille, R.S.~Freeland, D.J.~Heinzen, F.A.~van Abeelen,
        and B.J.~Verhaar,
%``Observation of a Feshbach Resonance in Cold Atom Scattering,''
Phys.\ Rev.\ Lett.\ {\bf 81}, 69 (1998).

\bibitem{Roberts98}
J.L.~Roberts {\it et al},
%J.L.~Roberts, N.R.~Claussen, J.P.~Burke, Jr., C.H.~Greene,
%        E.A.~Cornell, and C.E.~Wieman,
%``Resonant Magnetic Field Control of Elastic Scattering in Cold {$^{85}$Rb}''
 Phys.\ Rev.\ Lett. {\bf 81}, 5109 (1998).


%%%% atom-molecule resonance %%%%

\bibitem{Donley-nature2002-417}
E.A.~Donley {\it et al.}, Nature {\bf 417}, 529 (2002).

\bibitem{Cla02}
N.R.~Claussen {\it et al.},
%``Microscopic Dynamics in a Strongly Interacting Bose-Einstein Condensate''
Phys.\ Rev.\ Lett.\ {\bf 89}, 010401 (2002).

\bibitem{Rb-85-data}
N.R.~Claussen {\it et al},
%``Very-high-precision bound-state spectroscopy near a {$^{85}$Rb} Feshbach resonance''
 Phys.\ Rev.\ A {\bf 67}, 060701 (2003).


%%%% mean-field Hamiltonian, RG eq. %%%%

\bibitem{RPW}
L.~Radzihovsky, J.~Park and P.~Weichman,
 Phys.\ Rev.\ Lett. {\bf 92}, 160402 (2004).

\bibitem{RDSS}
M.W.J.~Romans, R.A.~Duine, S.~Sachdev and H.T.C.~Stoof,
 Phys.\ Rev.\ Lett. {\bf 93}, 020405 (2004).

\bibitem{LL}
Y.-W.~Lee, Y.-L.~Lee,
 Phys.\ Rev.\ B {\bf 70}, 224506 (2004).

\bibitem{BM}
S.~Basu, E.~Mueller, [arXiv:cond-mat/0507460].
%{\it Stability of Bosonic atomic and molecular condensates near a Feshbach resonance}


%%%% discrete scaling, Efimov phys %%%%

\bibitem{Braaten:2004rn}
E.~Braaten, H.-W.~Hammer,
 Phys.\ Rep. {\bf 428}, 259 (2006).

\bibitem{Efimov:1970}
V.~Efimov,
 Phys.\ Lett. {\bf 33B}, 563 (1970).

\bibitem{Efimov:1971}
V.~Efimov,
 Sov.\ J.\ Nucl.\ Phys.\ {\bf 12}, 589 (1971).

\bibitem{Efimov:1979}
V.~Efimov,
 Sov.\ J.\ Nucl.\ Phys.\ {\bf 29}, 546 (1979).

\bibitem{BHK99}
P.F.~Bedaque, H.-W.~Hammer, and U.~van~Kolck,
% ``Renormalization of the three-body system with short-range interactions,''
Phys.\ Rev.\ Lett.\ {\bf 82}, 463 (1999).
%        [arXiv:nucl-th/9809025].

\bibitem{BHK99b}
P.F.~Bedaque, H.-W.~Hammer, and U.~van~Kolck,
% ``The Three-Boson System with Short-Range Interactions,''
Nucl.\ Phys.\ A {\bf 646}, 444 (1999).
%        [arXiv:nucl-th/9811046].

\bibitem{Platter:2004qn}
L.~Platter, H.-W.~Hammer, and U.-G.~Mei{\ss}ner,
%``The Four-Boson System with Short-Range Interactions,''
Phys.\ Rev.\ A {\bf 70}, 052101 (2004).
%[arXiv:cond-mat/0404313].


%%%% Renormalization %%%%

\bibitem{Holland02}
S.J.J.M.F.~Kokkelmans {\it et al.},
 Phys.\ Rev.\ A {\bf 65}, 053617 (2002).


%%%% LSZ %%%%%

\bibitem{Peskin1995}
M.E.~Peskin, D.V.~Schroeder,
{\it An Introduction to Quantum Field Theory},
Addison-Wesley, (1995).


%%%% exact solution of 3-body %%%%

\bibitem{Sim81}
I.V.~Simenog and A.I.~Sitnichenko,
%``Effect of long-range interaction in a three-body system
%        with short-range forces,''
Doklady Academy of Sciences of the Ukrainian SSR (in Russian),
Ser.~A, {\bf 11}, 74 (1981).

\bibitem{MOG05}
J.H.~Macek, S.~Ovchinnikov, and G.~Gasaneo,
%``Exact solution for boson-diboson elastic scattering at zero energy
%       in the shape independent model,''
Phys.\ Rev.\ A {\bf 72}, 032709 (2005).


%%%% universality %%%%

\bibitem{Ho-2004-92}
T.-L.~Ho,
 Phys.\ Rev.\ Lett. {\bf 92}, 090402 (2004).

%\bibitem{Cowell-2002-88}
%S.~Cowell {\it et al.},
% Phys.\ Rev.\ Lett. {\bf 88}, 210403 (2002).


%%%% thesis %%%%%

\bibitem{zhang-thesis}
D.~Zhang, {\it Aspects of Cold Bosonic Atoms with A Large Scattering Length},
Ph.D thesis, The Ohio State University (2006).


\end{thebibliography}
\end{document}